\RequirePackage{fix-cm}
\documentclass[twocolumn,epjc3,superscriptaddress,unsortedaddress,floatfix]{svjour3}  
\smartqed  
\RequirePackage{graphicx}
\RequirePackage[colorlinks,citecolor=blue,urlcolor=blue,linkcolor=blue]{hyperref}
\RequirePackage{besphysics}
\RequirePackage{xcolor}
\RequirePackage{amsmath}
\RequirePackage{siunitx}
\RequirePackage{booktabs}
\RequirePackage{lineno}
\usepackage{cite}
\journalname{Eur. Phys. J. C}
\begin{document}

\title{\boldmath Improved measurement of the strong-phase difference $\dkp$ in quantum-correlated $D\bar{D}$ decays}

\date{Received: date / Accepted: date}

\authorrunning{BESIII Collaboration}
\author{{\center BESIII Collaboration\thanksref{e1}}\\ \\
M.~Ablikim$^{1}$, M.~N.~Achasov$^{11,b}$, P.~Adlarson$^{69}$, M.~Albrecht$^{4}$, R.~Aliberti$^{30}$, A.~Amoroso$^{68A,68C}$, M.~R.~An$^{34}$, Q.~An$^{65,52}$, X.~H.~Bai$^{60}$, Y.~Bai$^{51}$
% ...
}

\thankstext{e1}{e-mail: besiii-publications@ihep.ac.cn}

\institute{ Institute of High Energy Physics, Beijing 100049, People's Republic of China \\
$^{2}$ Beihang University, Beijing 100191, People's Republic of China\\
$^{3}$ Beijing Institute of Petrochemical Technology, Beijing 102617, People's Republic of China\\
$^{4}$ Bochum Ruhr-University, D-44780 Bochum, Germany\\
$^{5}$ Carnegie Mellon University, Pittsburgh, Pennsylvania 15213, USA\\
$^{6}$ Central China Normal University, Wuhan 430079, People's Republic of China\\
$^{7}$ Central South University, Changsha 410083, People's Republic of China\\
$^{8}$ China Center of Advanced Science and Technology, Beijing 100190, People's Republic of China\\
$^{9}$ COMSATS University Islamabad, Lahore Campus, Defence Road, Off Raiwind Road, 54000 Lahore, Pakistan\\
$^{10}$ Fudan University, Shanghai 200433, People's Republic of China\\
$^{11}$ G.I. Budker Institute of Nuclear Physics SB RAS (BINP), Novosibirsk 630090, Russia\\
$^{12}$ Guangxi Normal University, Guilin 541004, People's Republic of China\\
$^{13}$ Guangxi University, Nanning 530004, People's Republic of China\\
$^{14}$ Hangzhou Normal University, Hangzhou 310036, People's Republic of China\\
$^{15}$ Hebei University, Baoding 071002, People's Republic of China\\
$^{16}$ Helmholtz Institute Mainz, Staudinger Weg 18, D-55099 Mainz, Germany\\
$^{17}$ Henan Normal University, Xinxiang 453007, People's Republic of China\\
$^{18}$ Henan University of Science and Technology, Luoyang 471003, People's Republic of China\\
$^{19}$ Henan University of Technology, Zhengzhou 450001, People's Republic of China\\
$^{20}$ Huangshan College, Huangshan 245000, People's Republic of China\\
$^{21}$ Hunan Normal University, Changsha 410081, People's Republic of China\\
$^{22}$ Hunan University, Changsha 410082, People's Republic of China\\
$^{23}$ Indian Institute of Technology Madras, Chennai 600036, India\\
$^{24}$ Indiana University, Bloomington, Indiana 47405, USA\\
$^{25}$ INFN Laboratori Nazionali di Frascati , (A)INFN Laboratori Nazionali di Frascati, I-00044, Frascati, Italy; (B)INFN Sezione di Perugia, I-06100, Perugia, Italy; (C)University of Perugia, I-06100, Perugia, Italy\\
$^{26}$ INFN Sezione di Ferrara, (A)INFN Sezione di Ferrara, I-44122, Ferrara, Italy; (B)University of Ferrara, I-44122, Ferrara, Italy\\
$^{27}$ Institute of Modern Physics, Lanzhou 730000, People's Republic of China\\
$^{28}$ Institute of Physics and Technology, Peace Ave. 54B, Ulaanbaatar 13330, Mongolia\\
$^{29}$ Jilin University, Changchun 130012, People's Republic of China\\
$^{30}$ Johannes Gutenberg University of Mainz, Johann-Joachim-Becher-Weg 45, D-55099 Mainz, Germany\\
$^{31}$ Joint Institute for Nuclear Research, 141980 Dubna, Moscow region, Russia\\
$^{32}$ Justus-Liebig-Universitaet Giessen, II. Physikalisches Institut, Heinrich-Buff-Ring 16, D-35392 Giessen, Germany\\
$^{33}$ Lanzhou University, Lanzhou 730000, People's Republic of China\\
$^{34}$ Liaoning Normal University, Dalian 116029, People's Republic of China\\
$^{35}$ Liaoning University, Shenyang 110036, People's Republic of China\\
$^{36}$ Nanjing Normal University, Nanjing 210023, People's Republic of China\\
$^{37}$ Nanjing University, Nanjing 210093, People's Republic of China\\
$^{38}$ Nankai University, Tianjin 300071, People's Republic of China\\
$^{39}$ National Centre for Nuclear Research, Warsaw 02-093, Poland\\
$^{40}$ North China Electric Power University, Beijing 102206, People's Republic of China\\
$^{41}$ Peking University, Beijing 100871, People's Republic of China\\
$^{42}$ Qufu Normal University, Qufu 273165, People's Republic of China\\
$^{43}$ Shandong Normal University, Jinan 250014, People's Republic of China\\
$^{44}$ Shandong University, Jinan 250100, People's Republic of China\\
$^{45}$ Shanghai Jiao Tong University, Shanghai 200240, People's Republic of China\\
$^{46}$ Shanxi Normal University, Linfen 041004, People's Republic of China\\
$^{47}$ Shanxi University, Taiyuan 030006, People's Republic of China\\
$^{48}$ Sichuan University, Chengdu 610064, People's Republic of China\\
$^{49}$ Soochow University, Suzhou 215006, People's Republic of China\\
$^{50}$ South China Normal University, Guangzhou 510006, People's Republic of China\\
$^{51}$ Southeast University, Nanjing 211100, People's Republic of China\\
$^{52}$ State Key Laboratory of Particle Detection and Electronics, Beijing 100049, Hefei 230026, People's Republic of China\\
$^{53}$ Sun Yat-Sen University, Guangzhou 510275, People's Republic of China\\
$^{54}$ Suranaree University of Technology, University Avenue 111, Nakhon Ratchasima 30000, Thailand\\
$^{55}$ Tsinghua University, Beijing 100084, People's Republic of China\\
$^{56}$ Turkish Accelerator Center Particle Factory Group, (A)Istinye University, 34010, Istanbul, Turkey; (B)Near East University, Nicosia, North Cyprus, Mersin 10, Turkey\\
$^{57}$ University of Chinese Academy of Sciences, Beijing 100049, People's Republic of China\\
$^{58}$ University of Groningen, NL-9747 AA Groningen, The Netherlands\\
$^{59}$ University of Hawaii, Honolulu, Hawaii 96822, USA\\
$^{60}$ University of Jinan, Jinan 250022, People's Republic of China\\
$^{61}$ University of Manchester, Oxford Road, Manchester, M13 9PL, United Kingdom\\
$^{62}$ University of Muenster, Wilhelm-Klemm-Str. 9, 48149 Muenster, Germany\\
$^{63}$ University of Oxford, Keble Rd, Oxford, UK OX13RH\\
$^{64}$ University of Science and Technology Liaoning, Anshan 114051, People's Republic of China\\
$^{65}$ University of Science and Technology of China, Hefei 230026, People's Republic of China\\
$^{66}$ University of South China, Hengyang 421001, People's Republic of China\\
$^{67}$ University of the Punjab, Lahore-54590, Pakistan\\
$^{68}$ University of Turin and INFN, (A)University of Turin, I-10125, Turin, Italy; (B)University of Eastern Piedmont, I-15121, Alessandria, Italy; (C)INFN, I-10125, Turin, Italy\\
$^{69}$ Uppsala University, Box 516, SE-75120 Uppsala, Sweden\\
$^{70}$ Wuhan University, Wuhan 430072, People's Republic of China\\
$^{71}$ Xinyang Normal University, Xinyang 464000, People's Republic of China\\
$^{72}$ Yunnan University, Kunming 650500, People's Republic of China\\
$^{73}$ Zhejiang University, Hangzhou 310027, People's Republic of China\\
$^{74}$ Zhengzhou University, Zhengzhou 450001, People's Republic of China\\ \\
$^{a}$ Also at the Moscow Institute of Physics and Technology, Moscow 141700, Russia\\
$^{b}$ Also at the Novosibirsk State University, Novosibirsk, 630090, Russia\\
$^{c}$ Also at the NRC "Kurchatov Institute", PNPI, 188300, Gatchina, Russia\\
$^{d}$ Also at Goethe University Frankfurt, 60323 Frankfurt am Main, Germany\\
$^{e}$ Also at Key Laboratory for Particle Physics, Astrophysics and Cosmology, Ministry of Education; Shanghai Key Laboratory for Particle Physics and Cosmology; Institute of Nuclear and Particle Physics, Shanghai 200240, People's Republic of China\\
$^{f}$ Also at Key Laboratory of Nuclear Physics and Ion-beam Application (MOE) and Institute of Modern Physics, Fudan University, Shanghai 200443, People's Republic of China\\
$^{g}$ Also at State Key Laboratory of Nuclear Physics and Technology, Peking University, Beijing 100871, People's Republic of China\\
$^{h}$ Also at School of Physics and Electronics, Hunan University, Changsha 410082, China\\
$^{i}$ Also at Guangdong Provincial Key Laboratory of Nuclear Science, Institute of Quantum Matter, South China Normal University, Guangzhou 510006, China\\
$^{j}$ Also at Frontiers Science Center for Rare Isotopes, Lanzhou University, Lanzhou 730000, People's Republic of China\\
$^{k}$ Also at Lanzhou Center for Theoretical Physics, Lanzhou University, Lanzhou 730000, People's Republic of China\\
$^{l}$ Also at the Department of Mathematical Sciences, IBA, Karachi , Pakistan\\}

\onecolumn
\maketitle
\twocolumn
\sloppy

\begin{abstract}
The decay $\Dmkp$ is studied in a sample of quantum-correlated $D\bar{D}$ pairs, based on a data set corresponding to an integrated luminosity of 2.93\,fb$^{-1}$ collected at the $\psi(3770)$ resonance by the BESIII experiment.  
The asymmetry between $C\!P$-odd and $C\!P$-even eigenstate decays into $K^-\pi^+$ is determined to be ${\cal A}_{K\pi} = 0.132 \pm 0.011 \pm 0.007$, where the first uncertainty is statistical and the second is systematic. This measurement is an update of an earlier study exploiting additional tagging modes, including several decay modes involving a $K^0_L$ meson. The branching fractions of the $K^0_L$ modes are determined as input to the analysis in a manner that is independent of any strong phase uncertainty. Using the predominantly $C\!P$-even  tag $D\to \pi^+\pi^-\pi^0$ and the ensemble of $C\!P$-odd eigenstate tags, the observable ${\cal A}_{K\pi}^{\pi\pi\pi^0}$  is measured to be $0.130 \pm 0.012 \pm 0.008$.  The two asymmetries are sensitive to $\rkp \cos \dkp$, where $\rkp$ and $\dkp$ are the ratio of amplitudes and phase difference, respectively, between the doubly Cabibbo-suppressed and Cabibbo-favoured decays.
In addition, events containing $D \to K^-\pi^+$  tagged by $D \to K^0_{S,L} \pi^+\pi^-$  are studied in bins of phase space of the three-body decays. This analysis has sensitivity to both $\rkp \cos\dkp$ and $\rkp \sin\dkp$.  A fit to ${\cal A}_{K\pi}$,  ${\cal A}_{K\pi}^{\pi\pi\pi^0}$ and the phase-space distribution of the $D \to K^0_{S,L} \pi^+\pi^-$ tags yields
$\dkp= \left( 187.6 {^{+8.9}_{-9.7}}{^{+5.4}_{-6.4}} \right)\si\degree$, where external constraints are applied for $\rkp$ and other relevant parameters.   This is the most precise measurement of $\dkp$ in quantum-correlated $D\bar{D}$ decays.
\end{abstract}

\section{Introduction}
\label{sec:intro}

The decay $\Dzkmp$ and its suppressed counterpart $\Dzkpp$ play an important role in flavour physics~\footnote{Charge conjugation is implicit throughout this paper.}.  In particular, precise studies of $D^0$-$\Dzbar$ oscillations have been performed by measuring the dependence of the ratio of the $\Dzkpp$ to $\Dzkmp$ decay rates on decay time~\cite{Aaij:2012nva,Aaij:2017urz,Aaltonen:2013pja}.  Furthermore, high sensitivity to the $C\!P$-violating weak phase $\gamma$ of the Cabibbo-Kobayashi-Maskawa Unitarity Triangle is  attainable
by measuring observables in the decay $B^- \to DK^-$, $D \to K^+\pi^-$, where $D$ signifies a superposition of $D^0$ and $\bar{D}^0$ states~\cite{ADS0,ADS}. Finally, observables associated with $\Dzkmp$ and $\Dzkpp$ serve as important benchmarks for attempts to understand whether the observed level of $C\!P$ violation in charm decays can be accommodated within the Standard Model~\cite{LHCb:2019hro}. All of these studies can benefit from improved knowledge of the parameters governing $D^0 \to K^\mp\pi^\pm$ decays, which is  obtainable from charm-threshold data collected by the BESIII experiment.

The magnitude of the ratio of the doubly Cabibbo-suppressed $D^0$-decay amplitude to the $\Dzbar$ Cabibbo-favoured amplitude, $\rkp$, and the strong-phase difference between them, $\dkp$, are defined by
\begin{equation}
\rkp \exp\left( -i\dkp \right) = \frac{\langle K^+\pi^- | {D^0} \rangle}{\langle K^+\pi^- | \Dzbar \rangle} \, ,
\label{eq:dkpdef}
\end{equation}
and are important parameters for describing any process that involves the $K^\mp\pi^\pm$ final state~\footnote{Equation~\ref{eq:dkpdef}, subsequent expressions, and strong-phase differences are given in the convention $C\!P |D^0\rangle = |\Dzbar\rangle$.  Note that Ref.~\cite{Ablikim:2014gvw} uses an alternative definition that leads to a $180\si\degree$ offset in the reported value of $\dkp$.}.
In such processes it is also necessary to account for the effects of $D^0$-$\bar{D}^0$ oscillations.  This phenomenon is governed by the parameters  $x=2(m_1 - m_2)/(\Gamma_1 + \Gamma_2)$ and $y=(\Gamma_1 - \Gamma_2)/(\Gamma_1 + \Gamma_2)$, where $m_{1,2}$ and $\Gamma_{1,2}$ are the mass eigenstates and their corresponding decay widths, respectively. Here $C\!P$ violation is neglected in both tree-mediated charm decays and oscillations, which is a good approximation~\cite{Amhis:2019ckw}.  When measuring $D^0$-$\bar{D}^0$ oscillations in $\Dzkpp$ and $\Dzkmp$ decays, the time-dependent decay rate is, at leading order, a function of $\rkp$ and the rotated parameter $y' = y\cos\dkp - x\sin\dkp$.  Hence, knowledge of $\rkp$ and $\dkp$ is required in order to determine the oscillation parameters.  Conversely, studies with other decays of observables sensitive to $x$ and $y$, and a measurement of the branching-fraction ratio $\br{}(\Dzkpp)/\br{}(\Dzkmp)$ can be used to perform an indirect determination of the strong-phase difference.  The current ensemble of charm measurements yields $\rkp = \left( 5.87 \pm 0.02\right)\times 10^{-2} $, $\dkp = \left(191.7^{+3.6}_{-3.8}\right)\si\degree$, $x=\left(4.07 \pm 0.44 \right)\times 10^{-3}$ and $y=\left(6.47 \pm 0.24\right)\times 10^{-3}$~\cite{Amhis:2019ckw}.

The $C\!P$ asymmetry in $B^- \to D K^-$, $D\to K^+ \pi^-$ 
decays has been measured by the LHCb collaboration to be ${\cal A}_{C\!P} = -0.451 \pm 0.026$, where the uncertainty includes both statistical and systematic contributions, but is dominated by the former~\cite{LHCb:2020hdx}.  This observable has the following dependence on the underlying physics parameters:
\begin{equation}
    {\cal A}_{C\!P} = \frac{2 r_B \rkp \sin (\delta_B + \dkp) \sin \gamma}{(r_B)^2 + (\rkp)^2 + 2 r_B \rkp \cos (\delta_B + \dkp) \cos \gamma}\, ,
\end{equation}
where $\gamma \approx 65\si\degree$, and $r_B \approx 0.1$ is the amplitude ratio  and $\delta_B \approx 130\si\degree$  the strong-phase difference associated with the $B$-meson decay~\cite{LHCb:2021dcr}. The least well known of the parameters in this expression is $\dkp$, for which the current precision available from charm observables alone induces an uncertainty on the predicted value of ${\cal A}_{C\!P}$ that is around three times larger than that of the experimental determination.     LHCb  has recently performed a global fit of all its measurements that are sensitive to $\gamma$ from $b$-hadron decays, including ${\cal A}_{C\!P}$,  together with the ensemble of its charm-mixing results~\cite{LHCb:2021dcr}.  This fit yields $\dkp = \left(190.0^{+4.2}_{-4.1}\right)\si\degree$, which is a significantly more precise value than that obtained from charm data alone. 
Hence, improved knowledge of $\dkp$ from the charm system is desirable to obtain maximum information on $\gamma$ from $b$-hadron data.

Since the discovery of $C\!P$ violation in charm decays by the LHCb collaboration in 2019~\cite{LHCb:2019hro}, much discussion has taken place on whether the observed non-zero value of $\Delta A_{C\!P}$, which is the difference in $C\!P$ asymmetries between the modes $D^0 \to K^+K^-$ and $D^0 \to \pi^+\pi^-$, can be accommodated within the Standard Model~\cite{Lenz:2020awd}.  One way to validate the Standard Model explanation would be to find a consistent picture of SU(3)-flavour-breaking effects from final-state interactions across the family of charmed meson decays to two pseudoscalars. This approach has been pursued in Ref.~\cite{Buccella:2019kpn}, which performs a fit to $\Delta A_{C\!P}$ and the measured branching fractions of two-body decays of $D^0$, $D^+$ and $D^+_s$ mesons.  An output of this exercise is the prediction that $\dkp = \left( 183 \pm 5.7 \right)\si\degree$.  Therefore a determination of this phase difference, with similar or better precision to the prediction, will provide an indirect test of whether the observed value of $\Delta A_{C\!P}$ is compatible with Standard Model expectations.

Measurements of $\dkp$ may be performed at charm threshold, which are complementary to the indirect determination that comes from $D^0$-$\bar{D}^0$-oscillation studies.  The two mesons produced through the process $e^+e^- \to \psi(3770) \to D\bar{D}$ exist in an anti-symmetric wavefunction.  If one meson is reconstructed in the signal decay $\Dmkp$ and the other is reconstructed in a so-called {\it tagging mode} that is not a flavour eigenstate, but rather a superposition of $D^0$ and $\Dzbar$, the signal meson will also be in a superposition of these two states and the overall decay rate will depend on the phase difference between them.

In this paper, charm-threshold data from the BESIII experiment are analysed to measure $\dkp$ by following the above strategy.  Two classes of tagging mode are exploited: $C\!P$ eigenstates (and quasi $C\!P$ eigenstates), which bring sensitivity to $\rkp \cos \dkp$, and the self-conjugate, multi-body decays $K^0_{S,L}\pi^+\pi^-$, which bring information on both $\rkp \cos \dkp$ and $\rkp \sin \dkp$.  As $\rkp$ is well known, the results of the two analyses may be used to determine $\dkp$.  
Both analyses make extensive use of studies performed for previous BESIII publications: 
\begin{itemize}
    \item The $C\!P$-eigenstate analysis is an update of a previous measurement~\cite{Ablikim:2014gvw}, which it augments with  additional decay modes to achieve increased sensitivity.  Information on the $C\!P$-tagged yields for the majority of modes is taken from  an earlier analysis~\cite{Ablikim:2021cqw}, while the yields for the tagging modes reconstructed in isolation, required for normalisation purposes, are measured and presented here. Several new tags are added to the measurement, including decays that have $K^0_L$ mesons in the final state. It is important that the branching fractions of these so-called $K^0_L X$ modes, which are necessary inputs to the analysis, are determined with methods that make no use of $D \to K^-\pi^+$ decays.  Hence this paper also reports branching-fraction measurements for these channels, performed in a manner that relies solely on other $C\!P$ eigenstates as tagging modes.
    
    \item  The yields of events containing both $D\to K^-\pi^+$  and  $D \to K^0_{S,L}\pi^+\pi^-$ decays have been measured by BESIII for the input they provide on the strong-phase variation over the phase space of the multi-body modes~\cite{Ablikim:2020yif,Ablikim:2020lpk}.
    Here, this procedure is inverted: the earlier measurement is re-performed with the $D \to K^-\pi^+$ inputs removed, and the resulting knowledge of the multi-body strong-phase variation and related parameters, and the measured yields of $D\to K^-\pi^+$ decays tagged by $D \to K^0_{S,L}\pi^+\pi^-$, are exploited together to gain sensitivity to $\rkp \cos \dkp$ and $\rkp \sin \dkp$.

\end{itemize}

This paper is organised as follows. The detector, data sets and simulation samples are outlined in Sec.~\ref{sec:detector}.
Section~\ref{sec:klx} presents the determination of the branching fractions of the $D \to K^0_L X$ modes, the results of which are used in the  measurement with $C\!P$-eigenstate tags in Sec.~\ref{sec:cptags}.  The measurement with $D \to K^0_{S,L}\pi^+\pi^-$ tags is described in Sec.~\ref{sec:SCMBtag}. In Sec.~\ref{sec:combination}, the measurements are combined to obtain a determination of $\dkp$.  A summary and outlook are presented in Sec.~\ref{sec:summary}.

\section{Detector, data sets, and simulation samples}
\label{sec:detector}

The data analysed were collected by the BESIII detector~\cite{Ablikim:2009aa} from symmetric $e^+e^-$ collisions 
provided by the BEPCII storage ring~\cite{Yu:2016cof} at a centre-of-mass energy of 3773\,MeV, and correspond to an integrated luminosity of 2.93\,${\rm fb}^{-1}$.
The cylindrical core of the BESIII detector covers 93\% of the full solid angle and consists of a helium-based
 multilayer drift chamber~(MDC), a plastic scintillator time-of-flight
system~(TOF), and a CsI(Tl) electromagnetic calorimeter~(EMC),
which are all enclosed in a superconducting solenoidal magnet
providing a 1.0\,T magnetic field. The solenoid is supported by an
octagonal flux-return yoke with resistive-plate-counter muon-identification modules interleaved with steel. 
The charged-particle momentum resolution at $1\,{\rm GeV}/c$ is
$0.5\%$, and the resolution of the ionisation loss d$E$/d$x$  is $6\%$ for electrons
from Bhabha scattering. The EMC measures photon energies with a
resolution of $2.5\%$ ($5\%$) at $1$\,GeV in the barrel (end-cap)
region. The time resolution in the TOF barrel region is 68\,ps, while
that in the end-cap region is 110\,ps.

Simulated data samples are produced with a {\sc
geant4}-based~\cite{geant4} Monte Carlo (MC) package, which
includes the geometric description of the BESIII detector and the
detector response. The simulation models the beam-energy spread and initial-state radiation (ISR) in the $e^+e^-$
annihilations with the generator {\sc
kkmc}~\cite{ref:kkmc}.
The inclusive MC sample, which is used to study background contributions and is an order of magnitude larger than the real data set, includes the production of $D^0\bar{D}^0$ and $D^+D^-$ 
pairs from decays of the $\psi(3770)$, decays of the $\psi(3770)$ to light hadrons or charmonia, 
the production of $J/\psi$ and $\psi(3686)$ states through ISR, and the
continuum processes incorporated in {\sc kkmc}~\cite{ref:kkmc}.
Additional large samples are generated for exclusive final states in order to determine signal efficiencies.
The known decay modes are modelled with {\sc
evtgen}~\cite{Lange:2001uf} using branching fractions reported by the
Particle Data Group (PDG)~\cite{pdg}, and the remaining unknown charmonium decays
are modelled with {\sc lundcharm}~\cite{PhysRevD.62.034003,YANGRui-Ling:61301}. Final-state radiation~(FSR)
from charged final-state particles is incorporated using 
{\sc photos}~\cite{RICHTERWAS1993163}. No attempt is made to implement quantum-coherence effects in the $\psi(3770)$ sample.

\section{Measurement of $D^0 \to K^0_LX$ branching fractions with $C\!P$-eigenstate tags}
\label{sec:klx}

The measurement of $\rkp \cos \dkp$ relies on the signal mode being reconstructed together with tagging $C\!P$ eigenstates in so-called {\it double-tag} events.  A valuable subset of these $C\!P$ eigenstates contains the decays $D \to K^0_L \pi^0$, $D \to K^0_L \omega$ and $D \to K^0_L \pi^0\pi^0$, collectively denoted as $K^0_L X$ final states  (the low level of $C\!P$ violation that exists in the neutral kaon system is neglected). In order to make use of these modes in the analysis, it is necessary to know their branching fractions, so that the observed yield of double-tag events can be compared to the expectation in the absence of quantum correlations.   The branching fraction of $D^0 \to K^0_L \pi^0$ was measured by the CLEO collaboration with a relative precision of $7\%$, using a selection of flavour-specific decays as tagging modes including $D \to K^-\pi^+$~\cite{He:2007aj}. CLEO also measured this branching fraction and those of $D \to K^0_L \omega$ and $D \to K^0_L \pi^0\pi^0$ with a wider selection of tags as outputs of a global analysis  focused on the determination of $\dkp$~\cite{Asner:2012xb}, however, these results are not included in the PDG. 
The branching fraction of $D^0 \to K^0_L \omega$ has recently been measured by BESIII with a relative precision of $3\%$, again using the flavour-specific tags including $D \to K^-\pi^+$~\cite{BESIII:2022xhe}.
It is desirable to determine these quantities not only with the best possible precision, but also in a manner that is independent of $D \to K^-\pi^+$, so that they can be used as uncorrelated inputs in the $\dkp$ analysis.   This goal is achieved by selecting events in which the $D \to K_L X$ decays are tagged by decays into other $C\!P$-eigenstate modes.

To illustrate the method, consider the $C\!P$-even case $D \to K^0_L \pi^0$.   
Let $N(K^0_L \pi^0|T_{-})$ be the yield of double-tag events, which is the number of  $D \to K^0_L \pi^0$
decays tagged by a mode $T_-$ that is a $C\!P$-odd  eigenstate, 
and $\epsilon(K^0_L \pi^0|T_{-})$  be the efficiency for reconstructing such events. Furthermore, let $N(T_-)$ be the number of {\it single-tag} events, which is the number of observed decays to the $C\!P$-odd eigenstate, with no requirements on the other charm-meson decay in the event, and $\epsilon(T_-)$ be the corresponding reconstruction efficiency.  Then the branching fraction of the $C\!P$-even charm eigenstate $D_{+}$ can be written as
\begin{equation}
    \br{}(D_{+} \to K^0_L\pi^0) = \frac{N(K^0_L\pi^0|T_{-})}{N(T_{-})}\cdot \frac{\epsilon(T_{-})}{\epsilon(K^0_L\pi^0|T_{-})}\, ,
\label{eq:cpbr}
\end{equation}
and the branching fraction of the flavour-eigenstate, which is a superposition of $C\!P$-even and $C\!P$-odd eigenstates, is half of this:
\begin{equation}
\br{}(D^0 \to K^0_L\pi^0) = 0.5 \times  \br{}(D_+ \to K^0_L\pi^0)\, .   
\end{equation}
In the case where several $C\!P$ tags are used, this  branching fraction is given by
\begin{equation}
   \br{}(D^0 \to K^0_L\pi^0) = \frac{\sum N(K^0_L\pi^0|T^i_{-})}{2\, \sum  N(T^i_{-})\, \epsilon(K^0_L\pi^0|T^i_{-})/\epsilon(T^i_{-})}\, ,
\label{eq:cpbr1}   
\end{equation}
where the sum runs over all tags with the symbols for each tag designated by the superscript $i$.  
The $C\!P$-odd tags that are used in the analysis are $D \to K^0_S \pi^0$, $D\to K^0_S \eta$, $D\to \ksetap$ and $D\to K^0_S \omega$. The $D$-meson decay products are reconstructed through the modes: $K^0_S \to \pi^+\pi^-$, $\pi^0 \to \gamma\gamma$, $\eta \to \gamma\gamma$ and $\pi^+\pi^-\pi^0$, $\eta' \to \pi^+\pi^-\gamma$ and $\pi^+\pi^-\eta$, with $\eta \to \gamma \gamma$, and $\omega \to \pi^+\pi^-\pi^0$. $C\!P$-violation and matter-interaction effects within the neutral-kaon system are not considered because their impact is negligible in comparison to the experimental sensitivity.

The same relation applies to the $C\!P$-even mode $D \to K^0_L\omega$, and an analogous one for the $C\!P$-odd decay $D \to K^0_L\pi^0\pi^0$ for which $C\!P$-even tagging modes, $T_+$, are employed.
The $C\!P$-even tags used in the branching-fraction analysis are $D \to K^+K^-$, $D \to \pi^+\pi^-$, $D \to K_S^0\pi^0\pi^0$ and $D \to \pi^+\pi^-\pi^0$.  The latter mode has a $C\!P$-odd impurity of around 3\% that must be corrected~\cite{Malde:2015mha}.

Charged tracks must satisfy $|\!\cos\theta| < 0.93$, where $\theta$ is the polar angle with respect to the 
direction of the positron beam. The distance of closest approach of the track to the interaction point is required to be less than $10\,$cm in the beam direction (or $20\,$cm for the daughters of $\ks$ candidates) and less than $1\,$cm in the plane perpendicular to the beam (no requirement for $\ks$ daughters).
The d$E$/d$x$ and time-of-flight measurements are used to calculate particle-identification (PID) probabilities for the pion 
and kaon hypotheses.  The track is labelled a kaon or pion candidate, depending on which PID probability is higher. 
 
Photon candidates are selected from showers deposited in the EMC, with energies larger than $25\,$MeV in the barrel ($|\!\cos{\theta}|<0.80$) or $50\,$MeV in the end cap ($0.86<|\!\cos{\theta}|<0.92$). 
In order to suppress beam background or electronic noise, the shower clusters are required to be within $[0, 700]\,$ns of the start time of the event. When forming $\pi^0$ ($\eta$) candidates from pairs of photons, one photon is required to lie in the barrel region, where the energy resolution is the best, and the invariant mass of the pair is required to be within  [0.115, 0.150] ([0.480, 0.580]) \gev/$c^{2}$.  To improve momentum resolution, a kinematic fit is performed, where the reconstructed $\pi^0$ $(\eta)$ mass is constrained to the known value~\cite{pdg} and the resulting four-vector is used in the subsequent analysis.  When building $\omega$ and $\eta \to \pi^+\pi^-\pi^0$ candidates, the invariant mass of the $\pi^+\pi^-\pi^0$ combination is required to lie within [0.750, 0.820]\,\gev/$c^{2}$ and
[0.530, 0.565]\,\gev/$c^{2}$, respectively.
The invariant mass of the $\pi^+\pi^-\gamma$ and $\pi^+\pi^-\eta$  system used to form the $\eta'$ candidate must fall within [0.940, 0.970]\,\gev/$c^{2}$ and [0.940, 0.976]\,\gev/$c^{2}$, respectively.

Candidate $\ks$ mesons are reconstructed from pairs of tracks with opposite charge, and with no PID requirements. A flight-significance criterion is imposed, in which the distance from the beam spot  to the decay vertex, normalised by the uncertainty on this quantity, is required to be greater than two.
In addition, a constrained vertex fit is performed for each candidate, retaining those with a resulting invariant mass within [0.487, 0.511]\,\gev/$c^{2}$. 
 
To suppress combinatorial background, the energy difference, $\Delta{E} = E_{D} - \sqrt{s}/2$ is required to be within $\pm 3\sigma_{\Delta E}$ around the $\Delta E$ peak, where $\sigma_{\Delta E}$ is the $\Delta E$ resolution and $E_{D}$ is the reconstructed energy of a $D$ candidate in the rest frame of the $e^+e^-$ collision.  Cosmic and Bhabha backgrounds in the tag modes $D \to K^+K^-$ and $\pi^+\pi^-$ are suppressed by demanding that the two charged tracks have a TOF time difference less than $5\,$ns and that neither track is identified as an electron or a muon.
The $\pi^+\pi^-$ vertex in the mode $D \to \pi^+\pi^-\pi^0$ must have a flight significance of less than two, in order to reject $D\to K^0_S\pi^0$ decays.

Events containing a $K^0_L$ meson cannot be fully reconstructed and so are selected using a missing-mass technique.  The tagging mode is reconstructed, and its momentum, $\mathbf{p_T}$, is measured in the centre-of-mass frame of the $e^+e^-$ collision. If more than one candidate is found, the one with the smallest value of $|\Delta E|$ per mode is retained. Then the total energy, $E_{X}$, and momentum, $\mathbf {p_{X}}$, of the charged particles and $\pi^0$ candidates not associated with the tagging mode are determined.  This information allows the missing-mass squared,
\begin{linenomath*}
\begin{equation}
   M_{\rm miss}^2 = (\sqrt{s}/2-E_{X})^2/c^4-|\mathbf{p_{T}}+\mathbf{p_{X}}|^2/c^2,
   \label{eq:mm2}
\end{equation}
\end{linenomath*}
to be calculated, which should peak at the squared mass of the $K^0_{L}$ meson for signal events.

Vetoes are applied to suppress specific backgrounds.    Events are rejected in the  selection of $D \to K^0_L \omega$ decays that contain $\eta$ candidates or unused $\pi^0$ candidates in order to suppress contamination from $D \to \eta \omega$ and $D \to \pi^0 \omega$, respectively.  Similarly, background from $D \to \pi^0\pi^0\pi^0$ is suppressed in the selection of $D \to K^0_L \pi^0\pi^0$ decays by discarding events with unused $\pi^0$ candidates.  In addition, events containing unused charged tracks are also rejected for all selections, which suppresses contamination involving $\eta$-meson decays, and combinatorial backgrounds particularly in the higher $M^2_{\rm miss}$ region.

\begin{figure*}
    \centering
    \includegraphics[width=.8\textwidth]{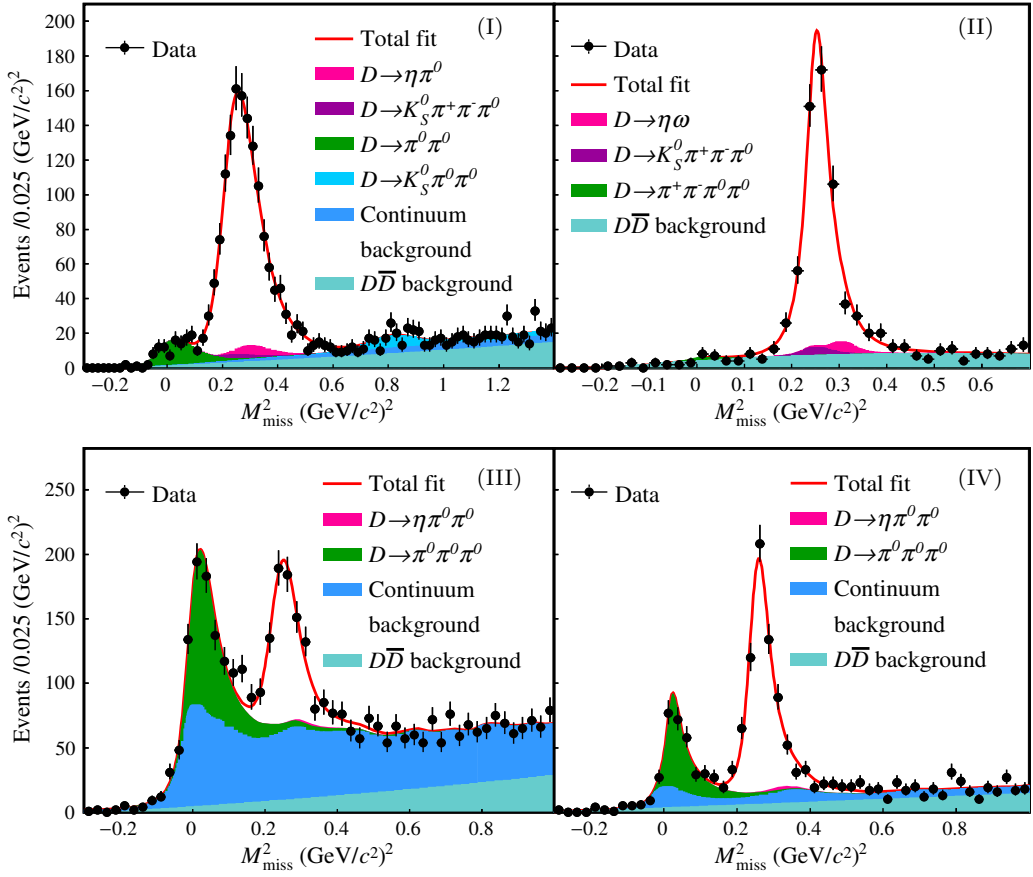}
    \put(-215,320){(I)}
    \put(-35,320){(II)}
    \put(-215,150){(III)}
    \put(-35,150){(IV)}
   \caption{Fits to the $M^2_{\rm miss}$ distributions of $D \to K_L^0 \pi^0$ candidates (I), $D \to K_L^0 \omega$ candidates (II), $D \to K_L^0 \pi^0 \pi^0$ candidates reconstructed together with $D \to \pi^+ \pi^- \pi^0$ tags (III) and with all other tags (IV).}
    \label{fig:klfit}
\end{figure*}

Figure~\ref{fig:klfit} shows the $M^2_{\rm miss}$ distribution for each of the signal decays integrated over all the tagging modes, apart from the case of $D \to K^0_L \pi^0\pi^0$ where the background level is significantly higher for  $D \to \pi^+\pi^-\pi^0$ tags, and hence is shown separately for these tags and for all other tags combined.  Clear signal peaks are observed around the squared mass of the $K^0_L$ meson, but background contributions are also visible or known to exist from studies of the MC simulation.  In the selection of $D \to K^0_L \pi^0$ there is contamination from $D \to \pi^0\pi^0$, $D \to \eta \pi^0$ and $D \to K^0_S \pi^0\pi^0$ decays, which occur at low, intermediate and high values of $M^2_{\rm miss}$, respectively.  In the case of $D \to K^0_L \omega$, there is background from $D \to \eta \omega$ decays at low  $M^2_{\rm miss}$, and a small contribution from $D \to \pi^+\pi^-\pi^0\pi^0$  decays. 
Both of these $C\!P$-odd signals have an approximately $2\%$ background that arises from non-resonant $D \to K^0_S \pi^+\pi^-\pi^0$ decays polluting the $D \to K^0_S \omega$ tags.
The most conspicuous peaking background in the $D \to K^0_L \pi^0 \pi^0$ analysis arises from $D \to \pi^0\pi^0\pi^0$ decays at low $M^2_{\rm miss}$, but there is also a contribution from $D \to \eta \pi^0 \pi^0$ under the signal.  For all selections there is a continuous spectrum of background that comes from $\psi(3770) \to D \bar{D}$ events and continuum $q\bar{q}$ production (apart from in the $D \to K^0_L \omega$ analysis, where this contribution is negligible). 

An unbinned maximum-likelihood fit is performed to determine the signal contribution for each of the distributions shown in Fig.~\ref{fig:klfit}. The range of each fit is the same as that of the individual plots, which differs from sample to sample on account of the different background sources.  The signal shape is modelled as a JohnsonSU function~\cite{JohsnsonSU}, with parameters determined from fits to MC samples, convolved with a Gaussian function to account for small differences in resolution between data and simulation.   The contributions of the $D\to\pi^0\pi^0$, $K_S^0\pi^0\pi^0$, $\pi^+\pi^-\pi^0\pi^0$ and $\pi^0\pi^0\pi^0$ backgrounds are also fitted, with their shapes described by appropriate functions fitted to MC simulation.  The non-peaking $\psi \to D\bar{D}$ background is modelled with a second-order polynominal with coefficients determined in the data fit.   The size and distribution of all other background components are taken from MC simulation, where in the case of $D \to \eta \pi^0$, $\eta \omega$ and $\eta \pi^0\pi^0$, the contributions are doubled to take account of the effect of quantum correlations, which are not included in the simulation. 

The fitted yield of $D \to K^0_L \omega$ events contains non-resonant $D \to K^0_L \pi^+\pi^-\pi^0$ background.  The size of this contribution is measured to be $(8.5 \pm 0.9)\%$ by studying the sidebands in either side of the $\omega$ peak in the $\pi^+\pi^-\pi^0$ invariant mass.  Fits are performed to the $M^2_{\rm miss}$ distribution in these regions and the results are interpolated within the $\omega$ mass window.

The measured signal yield of double-tagged $D \to K^0_L \pi^0$ and $D\to K^0_L \omega$ events is $1298 \pm 44$ and $494 \pm 27$, respectively.  About 60\% of these events are tagged with $D \to K^0_S \pi^0$ decays.  The measured signal yield of $D \to K^0_L \pi^0\pi^0$ events is $551 \pm 43$ when tagged by $D \to \pi^+\pi^-\pi^0$ decays and $644 \pm 31$ when selected with the other tags.  
The efficiencies of the double-tag selection are determined using dedicated MC samples and, by way of example, are found to be $(22.07 \pm  0.11)\%$ for $D \to K_L^0 \pi^0$ vs. $D \to K_S^0 \pi^0$, $(8.68\pm0.07)\%$ for $D \to K^0_L \omega$ vs. $D \to K_S^0 \pi^0$, and $(8.67 \pm 0.06)\%$ for $D \to K^0_L \pi^0\pi^0$ vs. $D \to \pi^+\pi^-\pi^0$ double tags, where daughter BFs are not included and the uncertainties are statistical. Information on the determination of the single-tag yields for the $C\!P$-eigenstates can be found in Sec.~\ref{sec:cptags}. These yields and the corresponding selection efficiencies are given in Table~\ref{tab:singletag}. Taking these inputs, and making use of Eq.~\ref{eq:cpbr1}, the branching fractions of the three signal modes are measured to be
\begin{eqnarray}
{\cal B}({D^0 \to K^0_L \pi^0}) & = & (0.97 \pm 0.03 \pm 0.02)\,\% \nonumber \\
{\cal B}({D^0 \to K^0_L \omega }) & = &  (1.09 \pm 0.06 \pm 0.03)\,\% \nonumber \\
{\cal B}({D^0 \to K_L^0 \pi^0\pi^0}) & = & (1.26 \pm 0.05 \pm 0.03)\,\% ,  \nonumber 
\end{eqnarray}
where the results have been corrected for the $\pi^0 \to \gamma \gamma$ and $\omega \to \pi^+\pi^-\pi^0$ branching fractions~\cite{pdg}.  The first uncertainty is statistical and the second systematic.  The results for 
$D^0 \to K^0_L \pi^0$ and  $D^0 \to K^0_L \omega$ are consistent with those obtained with flavour tags by CLEO and BESIII~\cite{He:2007aj,BESIII:2022xhe}.  The results for  $D^0 \to K^0_L \omega$ and $D^0 \to K_L^0 \pi^0\pi^0$ are around two and three sigma higher, respectively, than those reported in the CLEO global analysis~\cite{Asner:2012xb}, but are more precise.

The only sources of potential systematic bias in the measurement are associated with the yield determinations and the knowledge of the double-tag efficiencies.  All uncertainties related to the efficiency of the tag modes cancel in the ratio of double-tag to single-tag efficiencies in the denominator of Eq.~\ref{eq:cpbr1}. The assigned systematic uncertainties are summarised in Table~\ref{tab:systematics}. 

The uncertainties on the single-tag yields are listed in  Table~\ref{tab:singletag}, and are propagated to the branching-fraction measurement.  Uncertainties on the MC values for the individual charged-pion tracking and PID efficiencies, relevant for the $D \to K^0_ L \omega$ analysis are both assigned to be $0.5\%$~\cite{BESIII:2018apz}.  The uncertainty of the MC efficiency for reconstructing and identifying a neutral pion is set to be 1.0\%~\cite{BESIII:2018apz}.  All signal modes have a veto imposed for events with unused charged tracks, and subsets have a veto in place for events with unused $\pi^0$ candidate or an $\eta$ candidate.  Following Ref.~\cite{BESIII:2022xhe}, uncertainties of 1.0\%, 0.9\% and 0.1\% are assigned for each of these three conditions, reflecting the differences in efficiency between data and MC as measured in double-tagged $D \to K^0_S \pi^0$ events.   The uncertainty in the contamination from modes containing an $\eta$ meson ($\eta X$ background) is estimated by varying the contributions within one standard deviation of their measured branching fractions, and that of the non-resonant $\pi^+\pi^-\pi^0$ background in the $K^0_L \omega$ sample from propagating the statistical uncertainty in the fits to the sideband regions.  The parameters of the functions used to describe the signal have uncertainties from their fits to MC samples, which are propagated to the yield measurements.  In the case of the $D \to \pi^+\pi^-\pi^0$ tag a correction of 1/$F_+^{\pi\pi\pi^0}$ must be applied to the double-tag yield, where $F_+^{\pi\pi\pi^0} = 0.973 \pm 0.017$ is the measured $C\!P$-even fraction of this mode~\cite{Malde:2015mha}, thereby inducing a corresponding uncertainty in the yield measurement. Finally, the limitation in the knowledge of the double-tag efficiencies arising from the finite size of the MC samples contributes a small uncertainty.

\begin{table}[t]
\caption{Systematic uncertainties for the $D \to K_L^0 X$ branching fractions. 
For $D \to K_L^0\pi^0\pi^0$ the results are shown separately for events tagged with $D \to \pi^+\pi^-\pi^0$ decays and with other modes. Entries marked with `-' are where the source is not relevant. All uncertainties are relative and given in \%.}\label{tab:systematics}
\begin{center}
\begin{tabular}{lcccc}
\toprule
Source  &$K_L^0\pi^0$ &$K_L^0\omega$ &\multicolumn{2}{c}{$K_L^0\pi^0\pi^0$}\\
	& & & $\pi^+\pi^-\pi^0$ & Other \\
\midrule
Single-tag yields & 0.4 & 0.9 & 0.6 & 0.4\\
$\pi^{\pm}$ tracking &- &1.0 &-  &- \\
$\pi^{\pm}$ PID &- &1.0 &-	&- \\
$\pi^0$ reconstruction &1.0 &1.0 &2.0 &2.0\\
Track veto &1.0 &1.0 &1.0 &1.0\\
$\pi^0$ veto &- &0.9 &0.9 &0.9\\
$\eta$ veto &- &0.1 &- &- \\
$\eta X$ background &0.3 &0.4 &0.5 &0.5 \\
$\omega$ background &- &0.8 &- &-\\
Signal shape &0.9 &0.7 &1.6 &2.0 \\
$F_+^{\pi\pi\pi^0}$ &- &- & 1.7 & - \\
MC sample size &0.3 &0.5 &0.6 &0.3\\ 
$\br{}(\omega\to \pi\pi\pi^0)$ &- &0.8 &- &-\\\midrule
Total & 1.8 & 2.8 & 3.6 & 3.2 \\
\bottomrule
\end{tabular}
\end{center}
\end{table}

Various robustness tests are conducted; these have been successfully passed and thus lead to no additional systematic uncertainty.  These include verifying that consistent results are obtained when comparing subsets of tagging modes, and establishing that no signal is observed when attempts are made to reconstruct events containing two tag decays of the same $C\!P$ eigenvalue.

\section{Measurement of $\Akp$ and  $\mathcal{A}_{K\pi}^{\pi\pi\pi^0}$}
\label{sec:cptags}

 Observables sensitive to $\dkp$ can be constructed from ratios of event yields of suitably chosen samples. 
Let $N(K^-\pi^+|T_{+})$ be the number of $\Dmkp$ decays tagged by a mode that is a fully reconstructed $C\!P$-even  eigenstate, and $\epsilon(K^-\pi^+|T_{+})$  be the efficiency for reconstructing such events.  Then the branching fraction of the $C\!P$-odd charm eigenstate $D_{-}$ can be written as 
\begin{equation}
    \br{}(D_{-} \to K^-\pi^+) = \frac{N(K^-\pi^+|T_{+})}{N(T_{+})}\cdot \frac{\epsilon(T_{+})}{\epsilon(K^-\pi^+|T_{+})}.
    \label{eq:Dbr}
\end{equation}
An analogous expression may be written for the branching fraction of the $C\!P$-even eigenstate $D_+$ when tagged by a $C\!P$-odd decay.
However, when the $C\!P$ tag involves a $K^0_L$ meson, the double-tagged events must be reconstructed by a missing-mass technique, and it is not possible to reconstruct a single-tag sample. In this case  knowledge of the branching fraction of the $C\!P$ eigenstate is required to interpret the yield of double tags.  For example, if the tag is $C\!P$ even then
\begin{equation}
    \br{}(D_{-} \to K^-\pi^+) = \frac{N(K^-\pi^+|T_{+})}{2N_{D\bar{D}}\br{} (D^0 \to T_+)}\cdot \frac{1}{\epsilon(K^-\pi^+|T_{+})}\, ,
    \label{eq:Dbr_partial}
\end{equation}
where $N_{D\bar{D}}=(10,597 \pm 28 \pm 98) \times 10^3$ is the number of neutral $D$-meson pairs produced in the data set~\cite{BESIII:2018iev}. 

The asymmetry of the effective branching fraction is defined as
\begin{equation}
    \Akp \equiv \frac {\br{}(D_- \to K^-\pi^+) - \br{}(D_+ \to K^-\pi^+)}{\br{}(D_- \to K^-\pi^+) + \br{}(D_+ \to K^-\pi^+)}\, ,
\end{equation}
which to ${\cal O}\left(x,y, (\rkp)^2 \right)$ has the following relationship to the physics parameters:
\begin{equation}
\Akp = \frac{-2 \rkp \cos \dkp + y}{1 + (\rkp)^2}\, .
\label{eq:akp}
\end{equation}
Thus a measurement of the asymmetry allows $\rkp \cos \dkp$ to be determined, provided that other inputs are used to constrain $y$ and $\rkp$.

An earlier BESIII analysis~\cite{Ablikim:2014gvw} exploited eight $C\!P$-eigenstate tags.
The modes $D \to K^0_S \eta'$, $D \to K^0_S \phi$, $K^0_L \pi^0$, $K^0_L \omega$ and $K^0_L \pi^0 \pi^0$ are now added. One of the original eight tags was $D \to \rho^0 \pi^0$, a sub-mode of the decay $D \to \pi^+\pi^-\pi^0$.  The recent determination of $F_+^{\pi\pi\pi^0}$~\cite{Malde:2015mha}, the $C\!P$-even fraction of the three-body final state, now allows the inclusive decay to be used instead, which benefits the precision of the measurement due to the higher yield.
Although $F_+^{\pi\pi\pi^0}$ is very close to unity, it is still necessary to account for the small $C\!P$-odd content of the decay. Therefore a second asymmetry is defined
\begin{equation}
\label{eq:akpphys2}
    \mathcal{A}_{K\pi}^{\pi\pi\pi^0} \equiv \frac {\br{}(D_X \to K^-\pi^+) - \br{}(D_+ \to K^-\pi^+)}{\br{}(D_X \to K^-\pi^+) + \br{}(D_+ \to K^-\pi^+)},
\end{equation}
where $D_X$ is the superposition of $D_-$ and $D_+$ mesons  tagged by $D\to\pi^+\pi^-\pi^0$. To ${\cal O}\left(x,y, (\rkp)^2 \right)$ the dependence of this second asymmetry on the physics parameters is:
\begin{equation}
\label{eq:akpphys3}
\mathcal{A}_{K\pi}^{\pi\pi\pi^0} = \frac{\left(-2 \rkp \cos \dkp + y\right)F^{\pi\pi\pi^0}_{+}}{1 + (\rkp)^2+(1-F^{\pi\pi\pi^0}_{+})\left(2 \rkp \cos \dkp + y\right)}.
\end{equation}
The two asymmetries are both constructed from $C\!P$-odd tagged data and therefore have correlated uncertainties. However, this correlation can be taken into account when both asymmetries are combined to determine $\rkp\cos\dkp$.

It is noteworthy that the $C\!P$ tag $\ksphi$ has potential $S$-wave contamination under the $\phi$ peak that would lead to the reconstructed decay not being fully $C\!P$ odd. In this case, due to the low yield and hence low impact on the overall analysis, the tag is treated as a perfect eigenstate and this assumption is investigated as part of the systematic studies.

A summary of the  tags employed in the determination of $\Akp$ and $\mathcal{A}_{K\pi}^{\pi\pi\pi^0}$ is given in Table~\ref{tab:akpimodes}.  There are three modes which were not considered in the ${\cal{B}}(D^0 \to K^0_L X)$ analysis because of their limited statistical power: $D^0\to K^0_S \phi$, $\pi^0\pi^0$ and $K^0_S\pi^0\pi^0$.  
The first of these decays is reconstructed via $\phi \to K^+K^-$ with the requirement that the invariant mass of the kaon-pair squared lies within 0.010~${\rm GeV}^2/c^4$ of the known $\phi$-mass squared.

\begin{table}[!ht]
\caption{Summary of $C\!P$-eigenstate tag modes used in the determination of $\Akp$ and $\mathcal{A}_{K\pi}^{\pi\pi\pi^0}$.  }
\label{tab:akpimodes}
\begin{center}
\begin{tabular}{lc}
\toprule
$C\!P$ even &$\kk$, $\pipi$, $\pi^0\pi^0$ \\
&   $\kspiopio$, $\klpio$, $\klomega$\\
& \\
Quasi $C\!P$ even & $\pipipio$ \\
& \\
$C\!P$ odd & $\kspio$, $\kseta$, $\ksetap$,  \\
& $\ksomega$, 
 $\ksphi$,
$\klpiopio$ \\
\bottomrule
\end{tabular}
\end{center}
\end{table}

\begin{figure*}
    \centering
    \includegraphics[width=.98\textwidth]{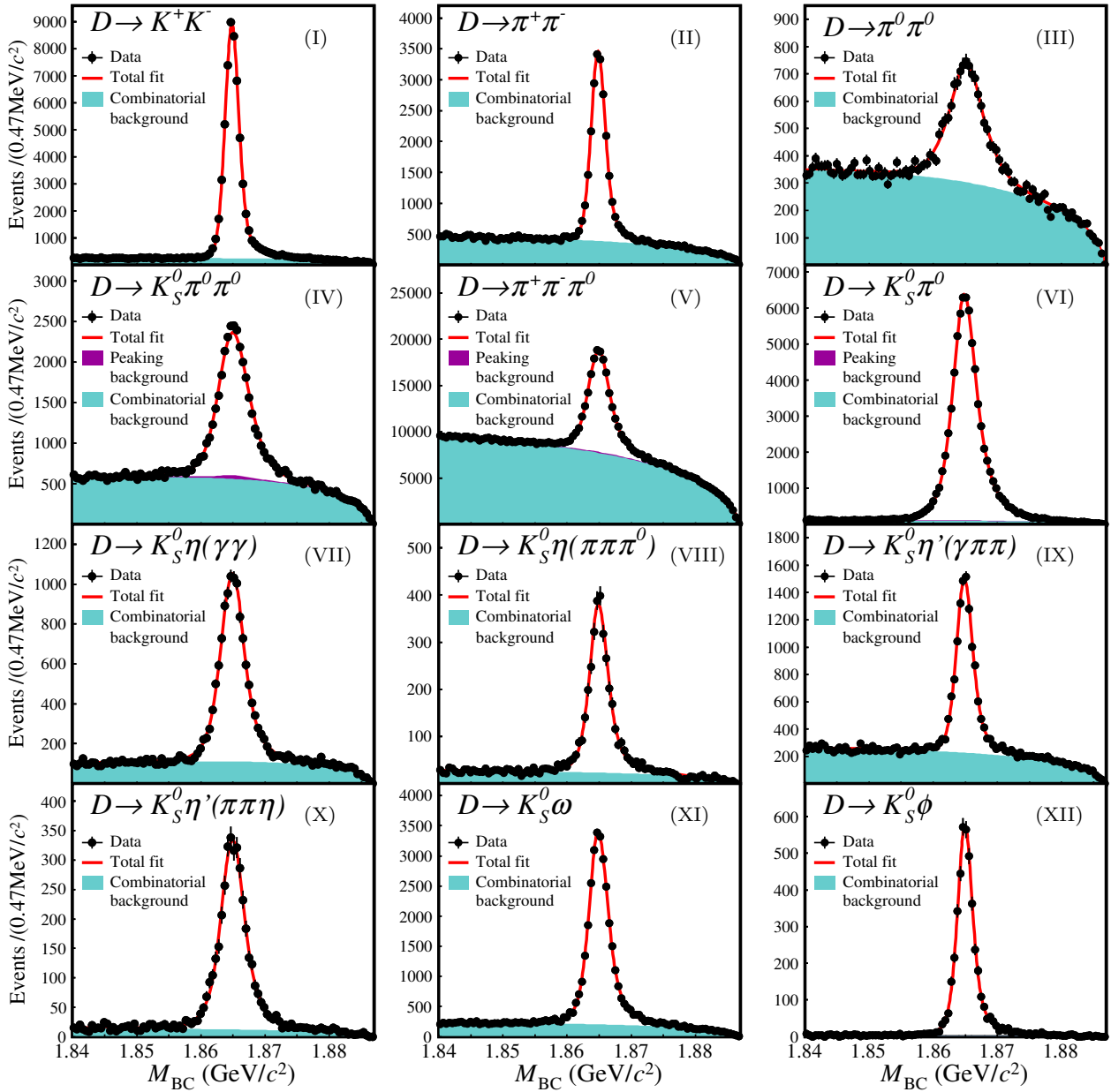}
    \put(-350,452){(I)}
    \put(-350,341){(IV)}
    \put(-350,231){(VII)}
    \put(-350,120){(X)}
    \put(-195,452){(II)}
    \put(-195,341){(V)}
    \put(-195,231){(VIII)}
    \put(-195,120){(XI)}
    \put(-40,452){(III)}
    \put(-40,341){(VI)}
    \put(-40,231){(IX)}
    \put(-40,120){(XII)}
   \caption{Fits to $M_{\rm BC}$ distributions of single-tag candidates for the $C\!P$-even (I-IV), quasi $C\!P$-even eigenstate (V) and $C\!P$-odd eigenstates (VI-XII). }
    \label{fig:CPsingletag_fit}
\end{figure*}

The single-tag yields  are determined through fits to the beam-constrained mass 
\begin{equation}
   M_{\rm BC} = \sqrt{(\sqrt{s}/2)^2/c^4-|{\mathbf{p_{D}}}|^2/c^2},
   \label{eq:mbc}
\end{equation}
for candidates lying within the $\pm 3 \sigma_{\Delta E}$ window.
Here $\mathbf{p_{D}}$ is the momentum of the $D$ candidate  in the rest frame of the $e^+e^-$ collision.
The fitted distributions are shown in Fig.~\ref{fig:CPsingletag_fit} for the $C\!P$-even and $C\!P$-odd modes. 
The signal shape is a template obtained from the corresponding signal MC, which is then convolved with a Gaussian function.   
The amount and shape of the peaking background contributions are taken from inclusive MC simulation. The peaking background is largest in the $D^0 \to K^0_S \pi^0\pi^0$ sample, where it is about 5\% of the signal.
For some modes this contribution is at a negligible level and is omitted in the fit.  
The shapes of the combinatorial background are described with an ARGUS function~\cite{ALBRECHT1990278}.  The single-tag yields from these fits are listed in Table~\ref{tab:singletag}, together with the efficiencies determined from MC simulation.  These results can be used to determine the branching fraction for each decay mode and are found to be compatible with those values reported in the PDG~\cite{pdg}.

\begin{table}[!ht]
\caption{Summary of single-tag yields.    Also shown are the selection efficiencies as determined from MC simulation, which are defined relative to the exclusive final state in which each channel is reconstructed. The uncertainties are statistical only.}
\label{tab:singletag}
\begin{center}
\begin{tabular}{lrr}
\toprule
Tag & Yield\hspace*{0.6cm} & Efficiency (\%)\hspace*{0.05cm} \\
\midrule
$K^+K^-$     &  55,696 $\pm$ 256 & 63.01 $\pm$ 0.05 \,\\
$\pi^+\pi^-$ &  20,403 $\pm$ 175 & 67.71 $\pm$ 0.08 \,\\
$\pi^0\pi^0$ &   7,012 $\pm$ 179 & 40.69 $\pm$ 0.12 \,\\
$\kspiopio$  &  29,328 $\pm$ 265 & 21.33 $\pm$ 0.04 \,\\
&& \\
$\pipipio$   & 129,601 $\pm$ 717 & 44.34 $\pm$ 0.02 \,\\
&& \\
$\kspio$                & 72,632 $\pm$ 294 & 40.50 $\pm$ 0.04 \,\\
$\kseta(\gamma\gamma)$  & 10,769 $\pm$ 131 & 36.11 $\pm$ 0.09 \, \\
$\kseta(\pi\pi\pi^0)$   &  3,054 $\pm$ $\,\,\,$67  & 17.76 $\pm$ 0.11 \,\\
$\ksetap(\gamma\pi\pi)$ & 10,427 $\pm$ 136 & 24.55 $\pm$ 0.07 \,\\
$\ksetap(\pi\pi\eta)$   &  3,723 $\pm$ $\,\,\,$70  & 15.27 $\pm$ 0.07 \,\\
$\ksomega$              & 25,794 $\pm$ 288 & 17.78 $\pm$ 0.03 \,\\
$\ksphi$                &  4,297 $\pm$ $\,\,\,$69  & 11.20 $\pm$ 0.06 \,\\
\bottomrule
\end{tabular}
\end{center}
\end{table}

\begin{figure*}
    \centering
    \includegraphics[width=.8\textwidth]{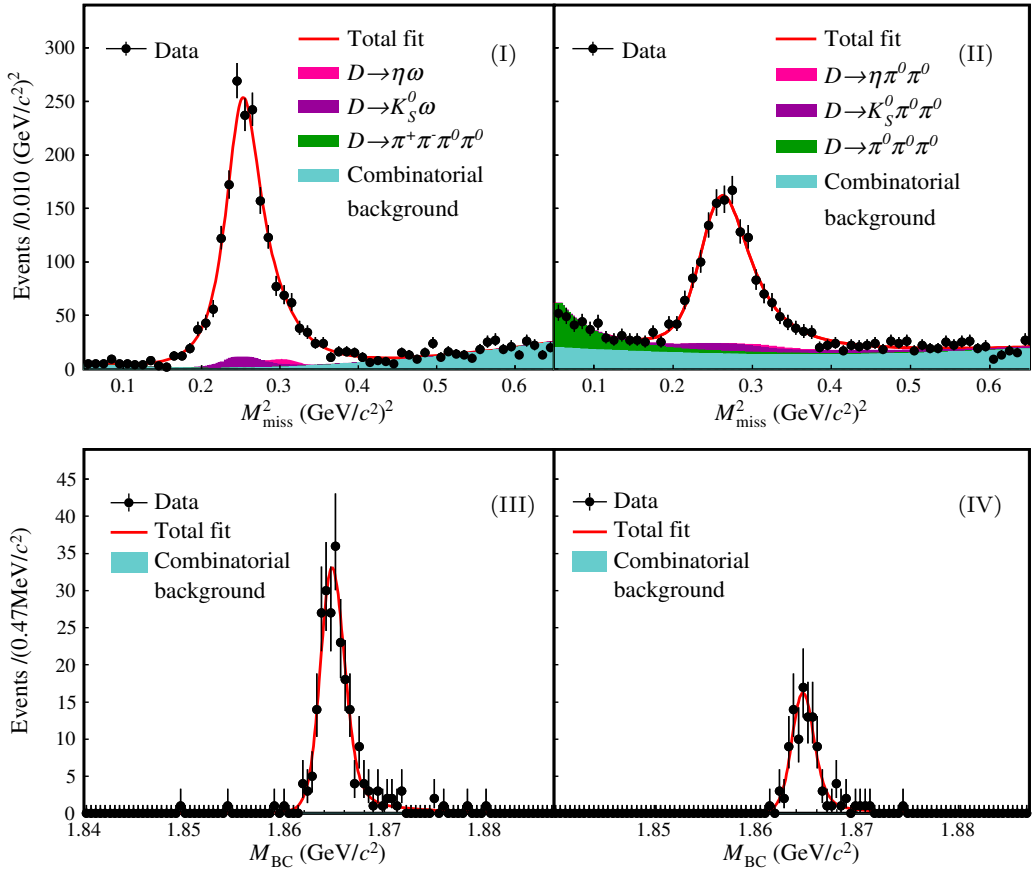}
    \put(-210,310){(I)}
    \put(-35,310){(II)}
    \put(-210,140){(III)}
    \put(-35,140){(IV)}
   \caption{Fits to $M_{\rm miss}^2$ distributions of double-tag candidates for the $D\to K_L^0\omega$ (I) and $D\to K_L^0\pi^0\pi^0$ (II) modes, and $M_{\rm BC}$ distributions of double-tag candidates for the $D\to\pi^0\pi^0$ (III) and $D\to K_S^0\phi$ (IV) modes. }
    \label{fig:DTCP}
\end{figure*}

Double-tag events containing both $D \to K^-\pi^+$ and a tag mode are selected. The yields of the fully reconstructed events are determined from a fit to the $M_{\rm BC}$ distribution on the tag side and those of the events containing a $D \to K^0_LX$ tag are obtained by fitting the $M^2_{\rm miss}$ distributions.
In the main the selection criteria, fit procedure and hence measured yields are identical to those reported in Ref.~\cite{Ablikim:2021cqw}, and so are not detailed here. Potential peaking backgrounds lying under the signal are estimated from MC simulation and, where necessary, corrected for quantum correlations. The only differences in selection are for the events tagged with $D \to K^0_L\omega$ and $D \to K^0_L\pi^0\pi^0$ where the requirements are adjusted to match those discussed in Sec.~\ref{sec:klx}, and for $D \to K^0_S \phi$ where the window imposed on the $K^+K^-$ invariant mass is made narrower to ensure the minimum level of $S$-wave contamination.  The sample of double tags containing $D \to \pi^0\pi^0$ decays was not selected in the analysis described in Ref.~\cite{Ablikim:2021cqw}, and is added for the current study.   The measured yields, and the selection efficiencies as determined from MC simulation, are presented in Table~\ref{tab:doubletag} and the fitted distributions for the new or updated double tags are shown in Fig.~\ref{fig:DTCP}.

\begin{table}[!ht]
\caption{Summary of double-tag yields. Also shown are the selection efficiencies as determined from MC simulation, which are defined relative to the exclusive final state in which each channel is reconstructed. The uncertainties are statistical only.}
\label{tab:doubletag}
\begin{center}
\begin{tabular}{lrr} \toprule
Tag & Yield\hspace*{0.3cm} & Efficiency (\%)\hspace*{0.2cm} \\
\midrule
$\kk$ 		 &1646 $\pm$ 42 & 43.21 $\pm$ 0.11\;\;\;\; \\	

$\pipi$		 & 592 $\pm$ 25 & 46.50 $\pm$ 0.11\;\;\;\;  \\

$\pi^0\pi^0$ 			& 235 $\pm$ 16 & 30.42 $\pm$ 0.10\;\;\;\; \\

$\kspiopio$  &804 $\pm$ 30	&	12.34 $\pm$ 0.07\;\;\;\; \\

$\klpio$ 	 &2590 $\pm$ 60	&25.05 $\pm$ 0.10\;\;\;\;  \\

$\klomega$ 	 &1357 $\pm$ 49 & 15.95 $\pm$ 0.07\;\;\;\;	\\

\\
 $\pipipio$ 	 &3647 $\pm$ 63 &28.20 $\pm$ 0.10\;\;\;\; \\

\\
$\kspio$ 	&1697 $\pm$ 42	& 28.36 $\pm$ 0.10\;\;\;\; \\

${\kseta}(\gamma\gamma)$  & 230 $\pm$ 16 &
24.97 $\pm$ 0.09\;\;\;\;  \\

${\kseta}(\pi\pi\pi^0)$    &66 $\pm$ \;\,9 &
13.04 $\pm$ 0.07\;\;\;\;  \\

${\ksetap}(\gamma\pi\pi)$ & 220 $\pm$ 16 &15.81 $\pm$ 0.07\;\;\;\; \\ 
${\ksetap}(\pi\pi\eta)$   &95 $\pm$ 10 &10.14 $\pm$ 0.06\;\;\;\; \\ 
${\ksomega}$ & 643 $\pm$ 28 & 12.07 $\pm$ 0.07\;\;\;\; \\

$\ksphi$ 				 &106 $\pm$ 10 & 7.11 $\pm$ 0.06\;\;\;\; \\

$\klpiopio$ 			& 1301 $\pm$ 54 & 12.96 $\pm$ 0.07\;\;\;\; \\  

\bottomrule
\end{tabular}
\end{center}
\end{table}

\begin{figure}
    \centering
    \includegraphics[width=.48\textwidth]{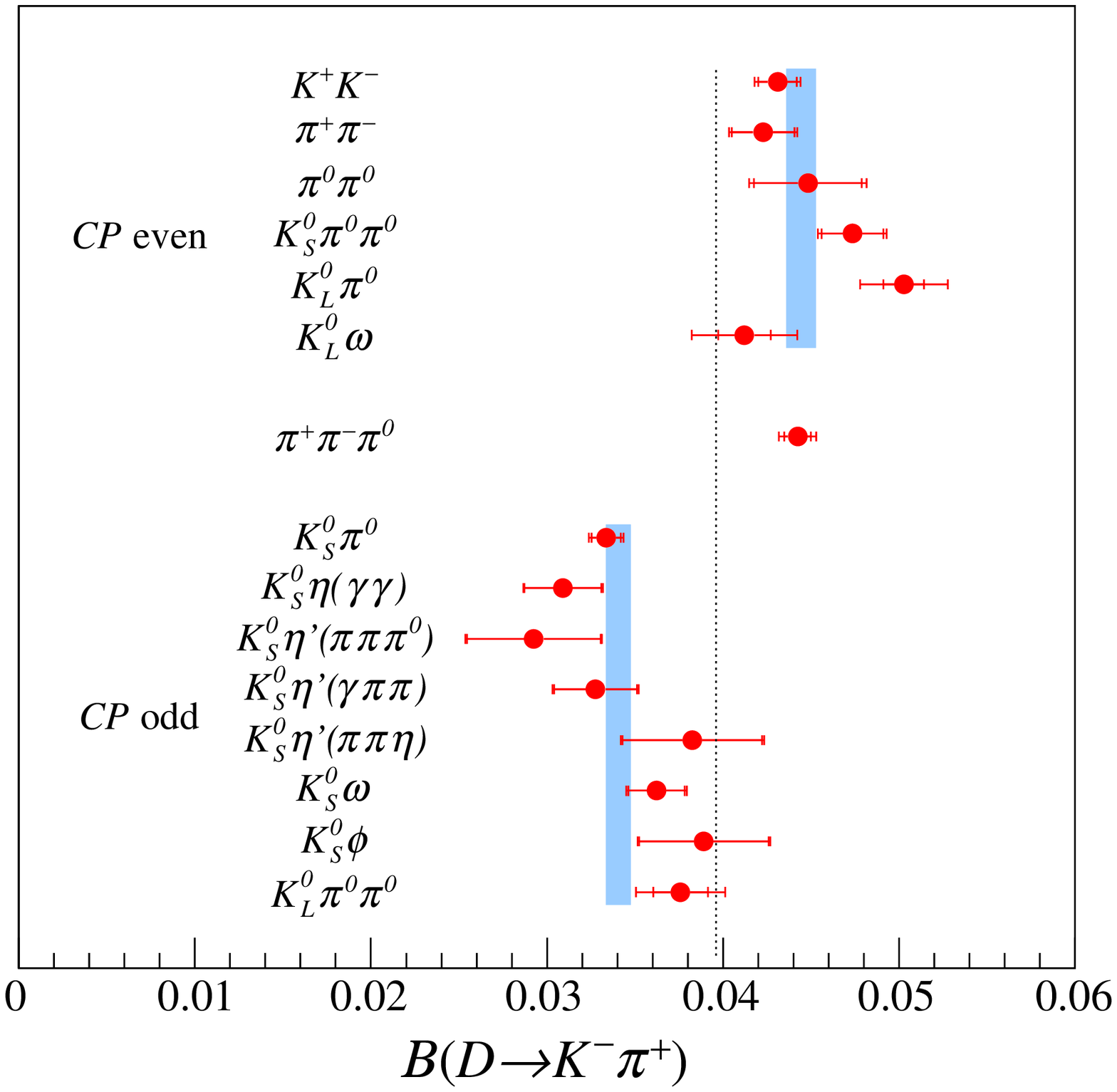}
    \caption{Branching fractions of $D_- \to K^-\pi^+$, $D_X \to K^-\pi^+$ and $D_+ \to K^-\pi^+$ determined with different tag modes. The outer error bars show the full uncertainty and the inner error bars the statistical uncertainty. The blue bands indicate the averaged result over each set of tags, and the vertical dotted line shows the measured central value of the $D^0 \to K^-\pi^+$ branching fraction~\cite{pdg}.}
    \label{fig:KpiBF}
\end{figure}

The $D_+$, $D_X$ and $D_- \to K^-\pi^+$ branching fractions are displayed in Fig.~\ref{fig:KpiBF} for each tag.  A least-squared fit is performed for the $C\!P$ eigenstates, taking account of the systematic uncertainties and their correlations, which yields ${\mathcal B}(D_- \to K^-\pi^+) =(4.445 \pm 0.060 \pm 0.056)\%$ with a fit quality per number of degrees of freedom (n.d.f.) of $\chi^2/{\rm n.d.f.} = 11.1/5$ and ${\mathcal B}(D_+ \to K^-\pi^+) =(3.406 \pm 0.059 \pm 0.038)\%$ with $\chi^2/{\rm n.d.f.} = 10.4/7$. Here the first uncertainty is statistical and the second systematic. The branching fraction obtained with the $D \to \pi^+\pi^-\pi^0$ tag is ${\mathcal B}(D_X \to K^-\pi^+) = (4.424 \pm 0.076 \pm 0.080)\%$, which, as expected, lies very close to the measurement of the $D_- \to K^-\pi^+$ branching fractions.  From these branching fractions it is found
\begin{eqnarray}
\Akp &=& 0.132 \pm 0.011 \pm 0.007 \nonumber \\
{\mathcal A_{K\pi}^{\pi\pi\pi^0}} &=& 0.130 \pm 0.012 \pm 0.008 \, , \nonumber
\end{eqnarray}
with correlation coefficients of 0.38 and 0.16 for the statistical and systematic uncertainties respectively. The result for $\Akp$ is consistent with that reported in Ref.~\cite{Ablikim:2014gvw} and is more precise.

In the determination of the branching fractions, the effects of several sources of possible systematic bias are evaluated, which are then propagated to the asymmetries.  The most significant of these arises from the knowledge of the $D \to K^0_L X$ branching fractions.  However, the uncertainties from this source that enter the determination of the $D_\pm \to K^-\pi^+$ branching fractions are significantly smaller than those reported in Sec.~\ref{sec:klx}, as many of the contributions considered in Table~\ref{tab:systematics} are common to both the $D \to K^0_L X$ branching fraction and the double-tag efficiency in the  denominator of Eq.~\ref{eq:Dbr_partial}, and thus cancel.
All double-tag efficiencies incur a relative uncertainty of 1\% associated with the knowledge of the reconstruction and identification efficiencies of the pion and kaon in the $D \to K^-\pi^+$ decay.   There are also uncertainties arising from the knowledge of $N_{D\bar{D}}$, the single-tag yields and the finite size of the MC samples used to determine the efficiencies.  The effect of possible $S$-wave contamination in the $D \to K^0_S \phi$ decay is studied, based on the results reported in Ref.~\cite{Libby:2010nu}, and is found to be negligible.  The systematic uncertainties on $\Akp$ and ${\mathcal A}^{\pi\pi\pi^0}_{K\pi}$ are summarised in Table~\ref{tab:asymsyst}.

\begingroup
\renewcommand{\arraystretch}{1.3} 
\begin{table}[!ht]
    \caption{Systematic uncertainties for $\Akp$ and  $\mathcal{A}_{K\pi}^{\pi\pi\pi^0}$.}
    \centering
    \begin{tabular}{lcc}\toprule
    Source &  $\Akp$ & $\mathcal{A}_{K\pi}^{\pi\pi\pi^0}$\\ \midrule
    ${\cal B}({D^0 \to K^0_L X})$ &0.0039  &0.0027\\
    Tracking and PID &0.0021 &0.0043\\
    $N_{D\bar{D}}$ &0.0014 &0.0010\\
    Single-tag yields &0.0040 &0.0049\\
    MC sample size &0.0039 &0.0043 \\ \midrule
    Total & 0.0072 & 0.0083  \\
    \bottomrule
    \end{tabular}
    \label{tab:asymsyst}
\end{table}
\endgroup

Using the measured asymmetries and external inputs for $y$, $(\rkp)^2$ and $F^{\pi\pi\pi^0}_+$~\cite{Amhis:2019ckw,Malde:2015mha}, it follows from Eqs.~\ref{eq:akp} and~\ref{eq:akpphys3} that
\begin{eqnarray}
    \rkp \cos\dkp = -0.0634 \pm 0.0048 \pm 0.0030 \pm 0.0004 \, ,\nonumber
\end{eqnarray}
where the final uncertainty arises from the knowledge of the external inputs.

\section{Measurement of $\rkp \cos\dkp$ and $\rkp \sin\dkp$ with $D \to K^0_{S,L} \pi^+\pi^-$  tags}
\label{sec:SCMBtag}

When the self-conjugate multi-body decay $D \to \kspipi$ is reconstructed as a tagging mode to $D \to K^-\pi^+$, the strong-phase variation over its Dalitz plot can be exploited to yield valuable information on $\dkp$.  This strong-phase variation has been  measured in studies at charm threshold by both the CLEO and BESIII collaborations~\cite{Libby:2010nu,Ablikim:2020yif,Ablikim:2020lpk}.

\begin{figure}
    \centering
    \includegraphics[width=.48\textwidth]{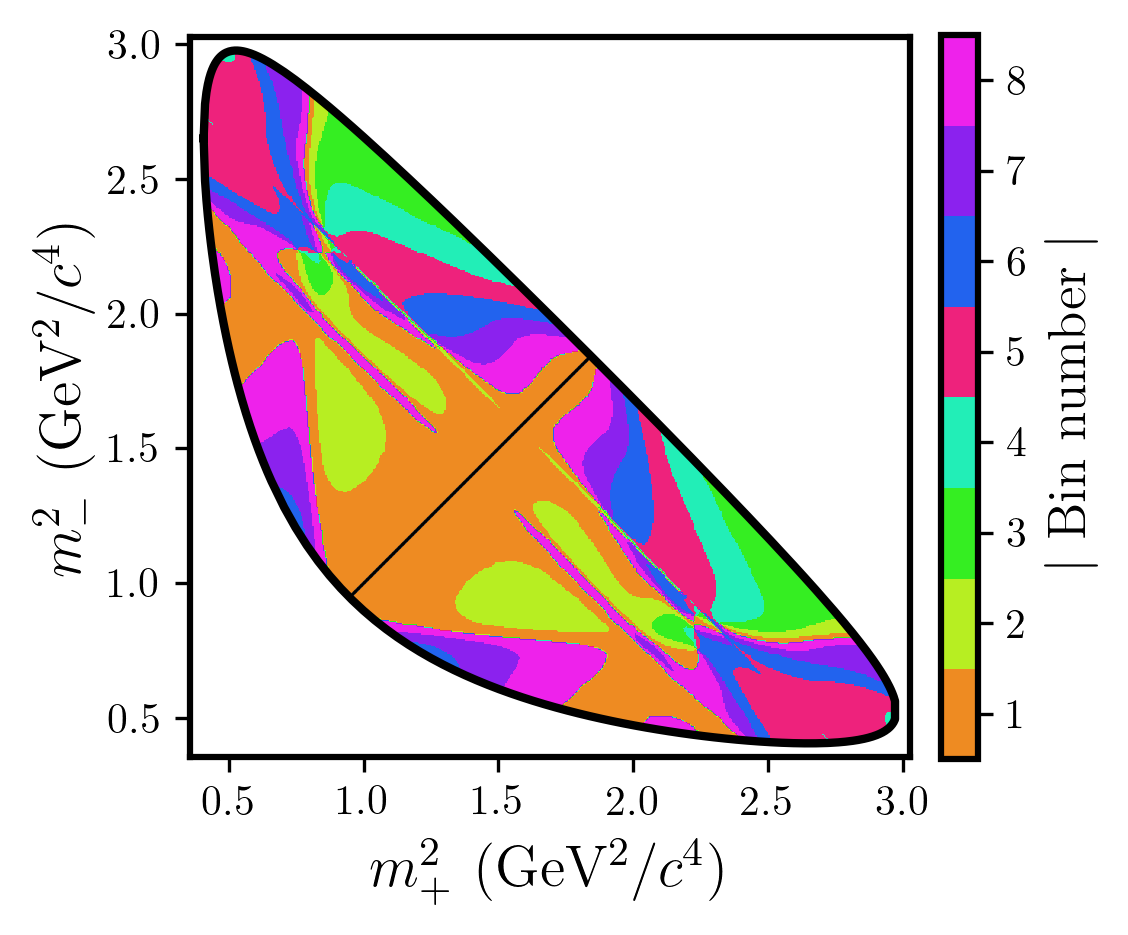}
    \caption{Dalitz plot of $D \to \kspipi$ decays, showing the equal-$\Delta \delta_D$ binning schemes.}
    \label{fig:kshhdalitz}
\end{figure}

The $D\to \kspipi$ Dalitz plot  has axes corresponding to the squared invariant masses $m_-^2 = m(\ks\pi^-)^2$ and $m_+^2= m(\ks \pi^+)^2$ for each $\ks$ and pion combination.
Eight pairs of bins are defined symmetrically about the line $m_-^2 = m_+^2$ such that 
the bin number changes sign under the exchange $(m_-^2,m_+^2) \leftrightarrow (m_+^2,m_-^2)$.  The bins are labelled $-8$ to 8 (excluding 0), with the positive bins lying in the region $m_+^2>m_-^2$. The strong-phase difference
between symmetric points in the Dalitz plot is given by $\Delta \delta_D^{\nkspipi} \equiv \delta_D^{\nkspipi}(m_+^2,m_-^2) - \delta_D^{\nkspipi}(m_-^2,m_+^2)$.  
The bin boundaries are chosen such that each bin spans an equal range in $\Delta \delta_D^{{\nkspipi}}$ (the so-called `equal-$\Delta \delta_D$ binning scheme'), as shown in Fig.~\ref{fig:kshhdalitz} where the variation in  $\Delta \delta_D^{\nkspipi}$ is assumed to follow that predicted by an amplitude model~\cite{Aubert:2008bd}. It is important to appreciate that though a model is used to define the bin boundaries, the values of $c_i$ and $s_i$ that are used come from direct measurements, and therefore cannot be biased through the choice of binning scheme.

Measurements performed with quantum-correlated $D\bar{D}$ pairs determine $c_i$, the cosine, and $s_i$, the sine of the strong-phase difference weighted by the $D^0$-decay amplitude ${ A}_{\nkspipi}(m_+^2,m_-^2)$ in bin $i$:
\begin{equation}
c_i = \frac{\int_{i}  {\rm d}{m_+^2} \, {\rm d}{m_-^2} \, A_{+} A_{-}\cos
\Delta \delta_D^{\nkspipi}({m_+^2},{m_-^2})   }
{\sqrt{\int_{i} {\rm d}{m_+^2}\, {\rm d}{m_-^2} \, (A_{+})^2 \,   \int_{i} {\rm d}{m_+^2} \, {\rm d}{m_-^2} \, (A_{-})^2    }}\, ,
\label{eq:ci}
\end{equation}
where
\begin{eqnarray}
 A_{+} = |{A}_{\nkspipi}(m_+^2,m_-^2)| & \,{\rm and} &  A_{-} = |{A}_{\nkspipi}(m_-^2,m_+^2)|,
 \nonumber
\end{eqnarray}
with an analogous expression for $s_i$.  Note that from these definitions it follows that $c_{-i} = c_{i}$ and $s_{-i}=-s_{i}$ in the absence of $C\!P$ violation.

When employing $D\to\kspipi$ as a tag mode, it is also necessary to know $K_i$, which is the probability of a single $D^0$ decay occurring in bin $i$: 
\begin{equation}
\label{eq:ki}
K_i = {\int_{i} {\rm d}{m_+^2} {\rm d}{m_-^2} (A_+)^2 }\bigg/{\sum_j \int_{j} {\rm d}{m_+^2} {\rm d}{m_-^2} (A_+)^2},
\end{equation}
where the sum in the denominator is over all bins.
This quantity may be measured in flavour-tagged decays.

Events in which one meson decays to $K^-\pi^+$ ($K^+ \pi^-$) and the other to $\ks\pi^+\pi^-$ are labelled with a negative (positive) bin number if $m_-^2<m_+^2$.
 Let $Y(K^-\pi^+|\kspipi)_{i}$ be the yield of double-tagged events in bin $i$ after correcting for any efficiency variation over the Dalitz plot.  Then it can be shown that~\cite{Ablikim:2021cqw}
\begin{multline}
\label{eq:kspipi}
Y(K^-\pi^+|\kspipi)_{i} =  H \left( K_{i} + \left( \rkp \right)^{2} K_{-i}\, -  \right. \\
\left. 2 \rkp \sqrt {K_{i} K_{-i}} \left[ c_{i} \cos\dkp - s_{i} \sin\dkp \right] \right)  ,
\end{multline}
where $H$ is a bin-independent normalisation factor. Hence a fit of $Y(K^-\pi^+|\kspipi)_{i}$  can be used to determine both $\rkp \cos \dkp$ and $\rkp \sin \dkp$.

Signal decays may also be tagged with the mode $D \to K^0_L \pi^+\pi^-$.  The same binning scheme is used, but the tag decay is now described by the parameters $c_i^\prime$, $s_i^\prime$ and $K_i^\prime$.  The yield of double-tagged events after correction for efficiency variation is given by
\begin{multline}
\label{eq:klpipi}
Y(K^-\pi^+|K^0_L\pi^+\pi^-)_{i} =  H^\prime \left( K^\prime_{i} + \left( \rkp \right)^{2} K^\prime_{-i}\, +  \right. \\
\left. 2 \rkp \sqrt {K^\prime_{i} K^\prime_{-i}} \left[ c^\prime_{i} \cos\dkp - s^\prime_{i} \sin\dkp \right] \right)  ,
\end{multline}
with $H^\prime$ the bin-independent normalisation factor for this tag.

    The $K_i^{(\prime)}$, $c_i^{(\prime)}$ and $s_i^{(\prime)}$ parameters have been measured by BESIII for $D \to K^0_{S,L} \pi^+\pi^-$ decays~\cite{Ablikim:2020yif,Ablikim:2020lpk}.  The $K_i^{(\prime)}$ parameters were determined by tagging the multi-body decays with the modes $D \to K^-\pi^+$, $D \to K^-\pi^+\pi^0$, $D \to K^-\pi^+\pi^+\pi^-$  and $D \to K^- e^+ \nu_e$  (for $D \to K^0_S \pi^+\pi^-$ only).  In order to interpret the hadronic decays  as pure flavour tags, it is necessary to correct their yields for the contribution of the doubly Cabibbo-suppressed amplitude.  In the case of $D \to K^-\pi^+$ this contribution manifests itself through the second two terms in Eqs.~\ref{eq:kspipi} and~\ref{eq:klpipi}, which carry the information on $\rkp$ and $\dkp$.   Therefore, for the current analysis, the $K_i^{(\prime)}$ parameters are re-determined without any $D \to K^-\pi^+$ inputs, by calculating a weighted average over the other flavour-tag results, and taking advantage of the most recent measurements of the hadronic parameters of the decays $D \to K^-\pi^+\pi^0$ and $D \to K^-\pi^+\pi^+\pi^-$, which are required to correct for the doubly Cabibbo-suppressed contamination in these modes~\cite{Ablikim:2021cqw}. The background estimations, which are around $3\%$ for $D \to \kspipi$ and $6\%$ for $D \to K^0_L \pi^+\pi^-$, are unchanged from the original analysis.  Also unchanged are the acceptance corrections, which vary by up to a relative $\pm 10\%$ per bin.  These corrections also account for migration effects between the bins, due to the finite invariant-mass resolution, which vary in the range of $(3-12)\%$ for $D \to \kspipi$ and $(3-18)\%$ for $D \to K^0_L \pi^+\pi^-$.
Table~\ref{tab:ki_kspipi} shows the re-calculated $K_i^{(\prime)}$ parameters and the $Y(K^-\pi^+|K^0_{S,L}\pi^+\pi^-)_i$ values following this procedure.  The latter numbers have been normalised to unity to allow for a convenient comparison with the $K_i^{(\prime)}$ values.

\begin{table*}[tb]
    \caption{Normalised yields of $D \to K^-\pi^+$ versus $D \to K^0_{S,L}\pi^+\pi^-$ double tags, corrected for acceptance effects in bins of the Dalitz plot. Also shown are $K_i^{(\prime)}$ values for $D \to K^0_{S,L}\pi^+\pi^-$, calculated with no $D \to K^-\pi^+$ inputs. The uncertainties for the yields are statistical, and in the case of $K_i^{(\prime)}$ include both the  statistical component and that from the doubly Cabibbo-suppressed correction. }\label{tab:ki_kspipi}
    \centering
    \begin{tabular}{r ll ll} \toprule
    Bin &  \multicolumn{1}{c}{$Y(K^-\pi^+$} & \multicolumn{1}{c}{$K_i$} & \multicolumn{1}{c}{$Y(K^-\pi^+$} & \multicolumn{1}{c}{$K_i^{\prime}$} \\
        &  \multicolumn{1}{c}{$|K^0_S \pi^+\pi^-)_i$} & &  \multicolumn{1}{c}{$|K^0_L \pi^+\pi^-)_i$} & \\
        \midrule
         1  & 0.1701 $\pm$ 0.0062 & 0.1780 $\pm$ 0.0033 & 0.1758 $\pm$ 0.0053 & 0.1859 $\pm$ 0.0033 \\
         2  & 0.0892 $\pm$ 0.0046 & 0.0873 $\pm$ 0.0024 & 0.0806 $\pm$ 0.0039 & 0.0789 $\pm$ 0.0023 \\
         3  & 0.0689 $\pm$ 0.0039 & 0.0668 $\pm$ 0.0021 & 0.0678 $\pm$ 0.0033 & 0.0628 $\pm$ 0.0021 \\
         4  & 0.0253 $\pm$ 0.0024 & 0.0232 $\pm$ 0.0013 & 0.0284 $\pm$ 0.0023 & 0.0224 $\pm$ 0.0013 \\
         5  & 0.0796 $\pm$ 0.0042 & 0.0847 $\pm$ 0.0024 & 0.0806 $\pm$ 0.0036 & 0.0728 $\pm$ 0.0021 \\
         6  & 0.0592 $\pm$ 0.0039 & 0.0567 $\pm$ 0.0021 & 0.0657 $\pm$ 0.0034 & 0.0620 $\pm$ 0.0020 \\
         7  & 0.1219 $\pm$ 0.0055 & 0.1261 $\pm$ 0.0029 & 0.1305 $\pm$ 0.0047 & 0.1255 $\pm$ 0.0027 \\
         8  & 0.1308 $\pm$ 0.0057 & 0.1347 $\pm$ 0.0030 & 0.1246 $\pm$ 0.0048 & 0.1363 $\pm$ 0.0030 \\
       $-1$ & 0.0973 $\pm$ 0.0046 & 0.0811 $\pm$ 0.0023 & 0.0824 $\pm$ 0.0035 & 0.0955 $\pm$ 0.0025 \\
       $-2$ & 0.0228 $\pm$ 0.0024 & 0.0189 $\pm$ 0.0011 & 0.0233 $\pm$ 0.0020 & 0.0218 $\pm$ 0.0013 \\
       $-3$ & 0.0220 $\pm$ 0.0022 & 0.0202 $\pm$ 0.0012 & 0.0203 $\pm$ 0.0020 & 0.0206 $\pm$ 0.0012 \\
       $-4$ & 0.0130 $\pm$ 0.0018 & 0.0160 $\pm$ 0.0011 & 0.0144 $\pm$ 0.0017 & 0.0128 $\pm$ 0.0010 \\
       $-5$ & 0.0452 $\pm$ 0.0032 & 0.0540 $\pm$ 0.0020 & 0.0433 $\pm$ 0.0028 & 0.0386 $\pm$ 0.0016 \\
       $-6$ & 0.0115 $\pm$ 0.0018 & 0.0121 $\pm$ 0.0010 & 0.0131 $\pm$ 0.0017 & 0.0100 $\pm$ 0.0009 \\
       $-7$ & 0.0118 $\pm$ 0.0018 & 0.0119 $\pm$ 0.0010 & 0.0169 $\pm$ 0.0019 & 0.0159 $\pm$ 0.0013 \\
       $-8$ & 0.0315 $\pm$ 0.0029 & 0.0284 $\pm$ 0.0015 & 0.0323 $\pm$ 0.0024 & 0.0381 $\pm$ 0.0017 \\
       \bottomrule
    \end{tabular}
\end{table*}

The $K_i^{(\prime)}$ values are also used as inputs in the determination of the $D \to K^0_{S,L}\pi^+\pi^-$ strong-phase parameters.  Therefore, it is desirable to re-calculate the values of $c_i^{(\prime)}$ and $s_i^{(\prime)}$ with the updated $K_i^{(\prime)}$ inputs.  The results are shown in Table~\ref{tab:cisi_kspipi}, and are found to be very similar to those reported in Refs.~\cite{Ablikim:2020lpk}. Furthermore, the differences in the correlation matrices between the two sets of results are negligible.  This behaviour is as expected, given the small weight that the $D \to K^-\pi^+$ inputs have in the original analysis.

\begin{table*}[ht]
    \caption{Values of $c_i^{(\prime)}$ and $s_i^{(\prime)}$ for $D \to K^0_{S,L} \pi^+\pi^-$, calculated with no $D \to K^-\pi^+$ inputs. For each value the first uncertainty is statistical and the second is systematic.} \label{tab:cisi_kspipi}
    \centering
    \begin{tabular}{lrrrr} \toprule
    Bin & \multicolumn{1}{c}{$c_i$} & \multicolumn{1}{c}{$s_i$} & \multicolumn{1}{c}{$c_i^\prime$} & \multicolumn{1}{c}{$s_i^\prime$} \\        \midrule
        1 &   0.708 $\pm$ 0.020 $\pm$ 0.009&    0.126 $\pm$ 0.076 $\pm$ 0.017& 0.796 $\pm$ 0.020 $\pm$ 0.013& 0.135 $\pm$ 0.078 $\pm$ 0.017\\
        2 &   0.676 $\pm$ 0.036 $\pm$ 0.019&    0.336 $\pm$ 0.134 $\pm$ 0.015& 0.854 $\pm$ 0.036 $\pm$ 0.018& 0.274 $\pm$ 0.137 $\pm$ 0.016\\
        3 &$-$0.002 $\pm$ 0.047 $\pm$ 0.018&    0.893 $\pm$ 0.113 $\pm$ 0.021& 0.174 $\pm$ 0.047 $\pm$ 0.016& 0.840 $\pm$ 0.118 $\pm$ 0.022\\
        4 &$-$0.601 $\pm$ 0.053 $\pm$ 0.017&    0.724 $\pm$ 0.142 $\pm$ 0.022&-0.501 $\pm$ 0.055 $\pm$ 0.019& 0.785 $\pm$ 0.146 $\pm$ 0.022\\
        5 &$-$0.964 $\pm$ 0.019 $\pm$ 0.013&    0.018 $\pm$ 0.081 $\pm$ 0.009&-0.972 $\pm$ 0.021 $\pm$ 0.017&-0.009 $\pm$ 0.089 $\pm$ 0.009\\
        6 &$-$0.561 $\pm$ 0.062 $\pm$ 0.025& $-$0.595 $\pm$ 0.147 $\pm$ 0.032&-0.392 $\pm$ 0.069 $\pm$ 0.026&-0.649 $\pm$ 0.153 $\pm$ 0.036\\
        7 &   0.044 $\pm$ 0.057 $\pm$ 0.023& $-$0.689 $\pm$ 0.143 $\pm$ 0.030& 0.465 $\pm$ 0.057 $\pm$ 0.019&-0.553 $\pm$ 0.160 $\pm$ 0.032\\
        8 &   0.398 $\pm$ 0.036 $\pm$ 0.017& $-$0.477 $\pm$ 0.091 $\pm$ 0.027& 0.631 $\pm$ 0.036 $\pm$ 0.016&-0.402 $\pm$ 0.099 $\pm$ 0.026\\ \bottomrule
    \end{tabular}
\end{table*}

\begin{figure}
    \centering
    \includegraphics[width=0.50\textwidth,trim={0 5.8cm 0 0},clip]{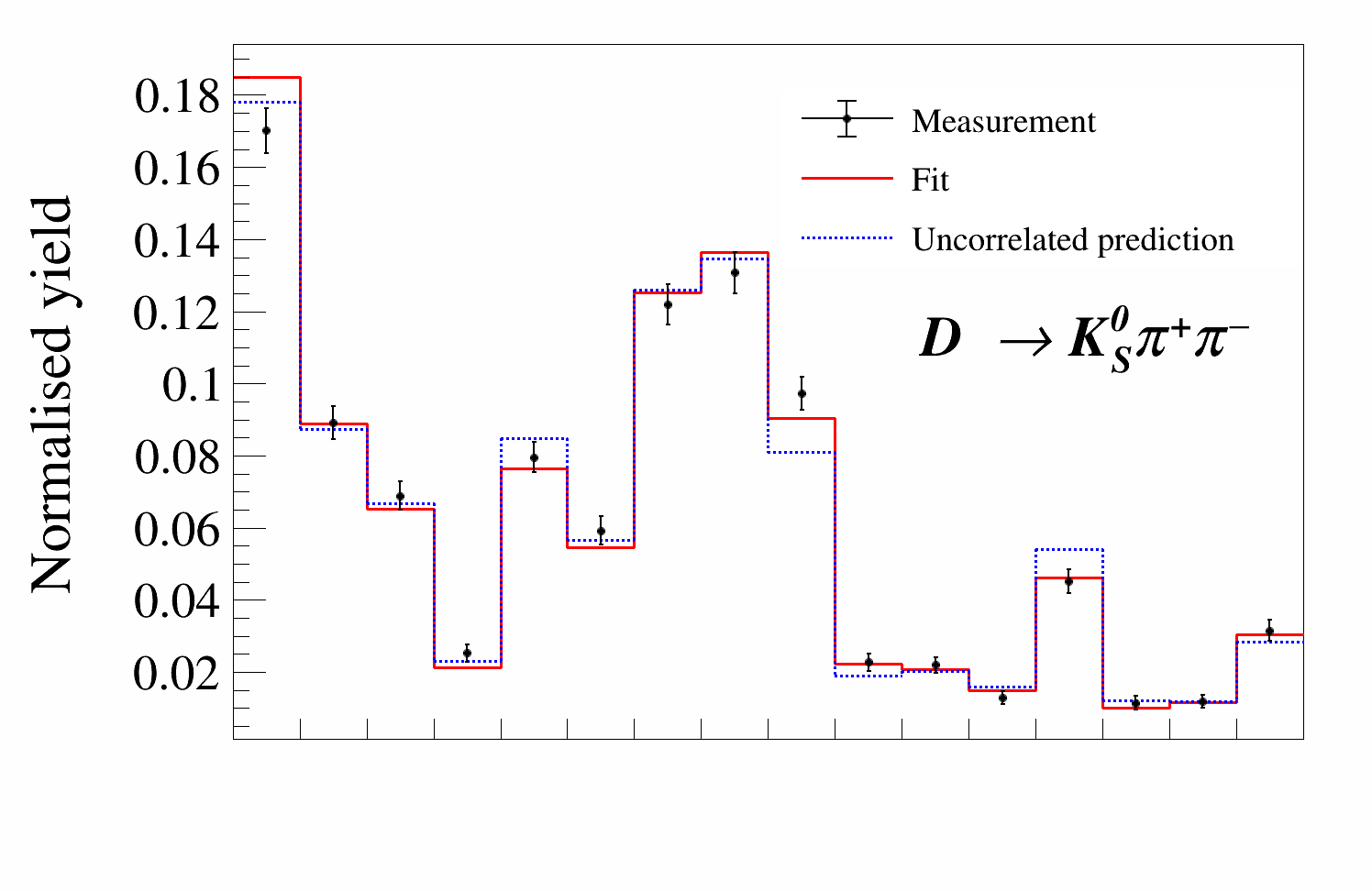}
    \includegraphics[width=0.50\textwidth,trim={0 0 0 1.73cm},clip]{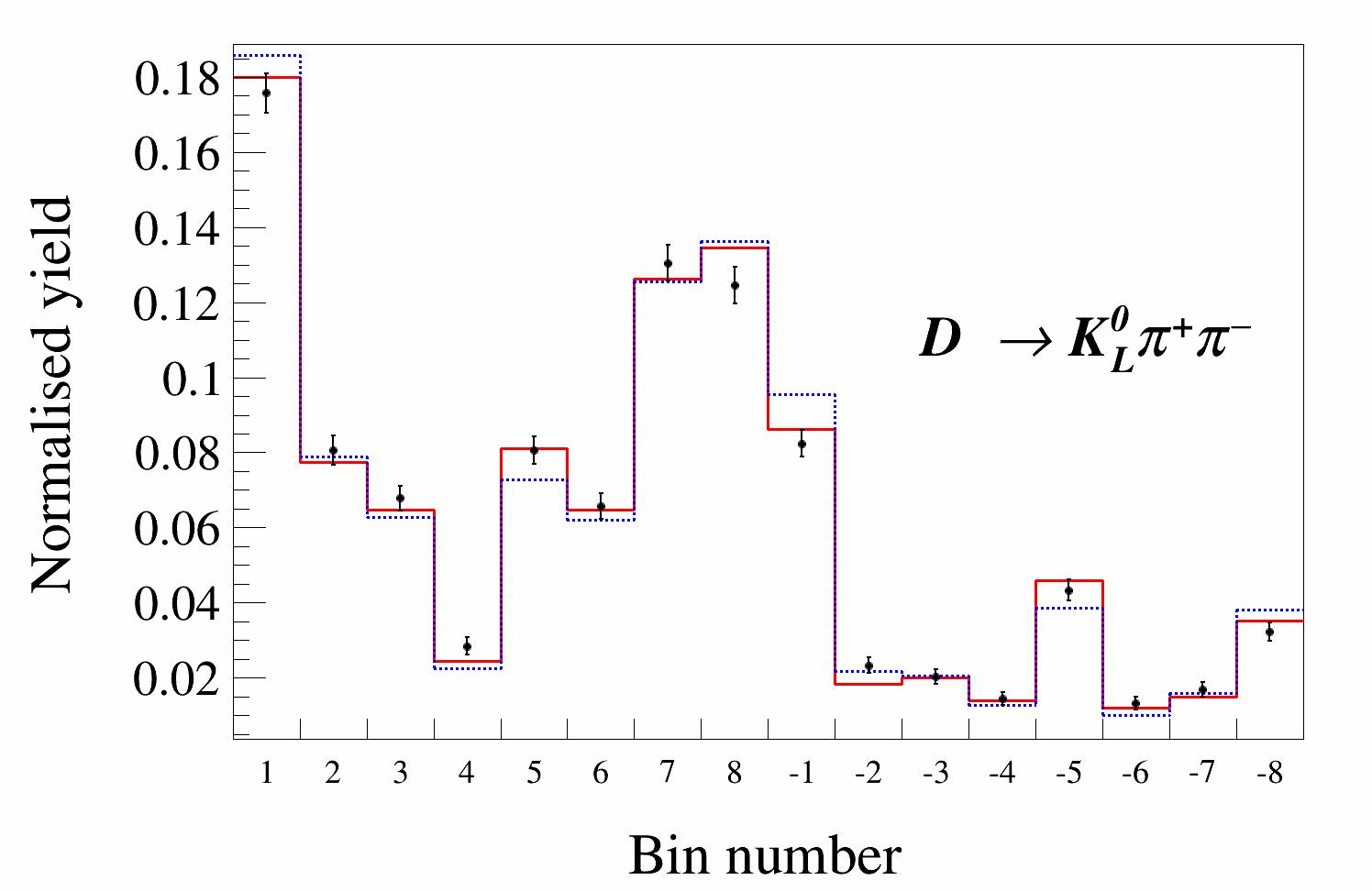}
    \caption{Fits to the $D \to K^-\pi^+$ sample tagged with $D \to K^0_S\pi^+\pi^-$  decays and $D \to K^0_L\pi^+\pi^-$ decays.}
    \label{fig:k0pipi_fit}
\end{figure}

A $\chi^2$ fit is performed to the normalised yields in the 32 phase-space bins of the two tagging modes, as listed in Table~\ref{tab:ki_kspipi}, with $\rkp\cos\dkp$ and $\rkp\sin\dkp$ as free parameters.
The expected yield values in the fit assume the distributions described by Eqs.~\ref{eq:kspipi} and~\ref{eq:klpipi} and use the values of $K_i^{(\prime)}$ from Table~\ref{tab:ki_kspipi}, and the values of $c_i^{(')}$ and $s_i^{(\prime)}$ from Table~\ref{tab:cisi_kspipi}.  The correlation matrices for $c_i^{(\prime)}$ and $s_i^{(\prime)}$  are taken from Ref.~\cite{Ablikim:2020lpk}.  The results and the $\chi^2/{\rm n.d.f.}$ are presented in Table~\ref{tab:kshh_fit} for the default fit for both tagging modes, as well as for separated fits to $D\to K^0_S\pi^+\pi^-$ and $D\to K^0_L\pi^+\pi^-$.  All fits are of good quality, and the two tags give compatible results. Figure~\ref{fig:k0pipi_fit} displays the fit to the full set of double tags.

\begin{table}[ht]
    \caption{Fit results for the $D \to K^-\pi^+$ decays tagged separately with $D \to K^0_S \pi^+\pi^-$, $D \to K^0_L \pi^+\pi^-$ and with both tags together.  
    The uncertainties are statistical only.}
    \label{tab:kshh_fit}
    \centering
    \begingroup
    \setlength{\tabcolsep}{4pt}
    \renewcommand{\arraystretch}{1.1}
    \begin{tabular}{lrrr} \toprule
    Sample & \multicolumn{1}{c}{$\rkp \cos\dkp$} & \multicolumn{1}{c}{$\rkp \sin \dkp$} & \multicolumn{1}{c}{$\chi^2/{n.d.f.}$} \\  \midrule
      $K^0_S \pi^+\pi^-$     & $-$0.0521 $\pm$ 0.0128 &   0.000 $\pm$ 0.017 & 16.5/14 \\
      $K^0_L \pi^+\pi^-$     & $-$0.0590 $\pm$ 0.0104 & $-$0.020 $\pm$ 0.015 & 21.1/14 \\
      $K^0_{S,L} \pi^+\pi^-$ & $-$0.0562 $\pm$ 0.0081 & $-$0.011 $\pm$ 0.012 & 38.6/30 \\ \bottomrule
    \end{tabular}
    \endgroup
\end{table}

The systematic uncertainties on the fit results come from two sources:  the uncertainties on the values of $K_i^{(\prime)}$, which consist of the statistical component listed in Table~\ref{tab:ki_kspipi} together with a significantly smaller contribution associated with the doubly Cabibbo-suppressed correction, and those on the values on $c_i^{(\prime)}$ and $s_i^{(\prime)}$ from Table~\ref{tab:cisi_kspipi}.  To quantify the effect of this imperfect knowledge, the fit is repeated many times with the values of these parameters randomly modified according to a Gaussian function of width set to the known uncertainty on each parameter, with correlations considered in the  $c_i^{(\prime)}$ and $s_i^{(\prime)}$ cases. The spread in the distribution of fit results is assigned as the systematic uncertainty.

The results, including the systematic uncertainties, are 
\begin{eqnarray}
    \rkp \cos\dkp &=& -0.0562 \pm 0.0081 \pm 0.0050 \pm 0.0010 \nonumber\\
    \rkp \sin\dkp &=& -0.011  \pm 0.012  \pm 0.007  \pm 0.003,
\nonumber
\end{eqnarray}
where the first uncertainties are statistical, the second are from the knowledge of the $K_i^{(\prime)}$ parameters and the third from the knowledge of the $c_i^{(\prime)}$ and $s_i^{(\prime)}$ parameters.  The correlation coefficient between the two results is 0.02. The measured value for $\rkp \cos \dkp$ is in good agreement with that obtained from the ${\cal A}_{K\pi}$ and ${\cal A}^{\pi\pi\pi^0}_{K\pi}$ measurements, reported in Sec.~\ref{sec:cptags}.

\section{Determination of $\dkp$}
\label{sec:combination}

The value of $\dkp$ is determined from a $\chi^2$ fit that uses the measurements of ${\cal A}_{K\pi}$ and ${\cal A}^{\pi\pi\pi^0}_{K\pi}$ as inputs, as well as the results for $\rkp \cos \dkp$ and $\rkp \sin \dkp$ obtained from the $D \to K^0_{S,L}\pi^+\pi^-$ analysis.  The dependencies of ${\cal A}_{K\pi}$, ${\cal A}^{\pi\pi\pi^0}_{K\pi}$ are taken from Eqs.~\ref{eq:akp} and~\ref{eq:akpphys3}, respectively. The auxiliary parameters $\rkp$, $y$ and $F^{\pi\pi\pi^0}_+$ are also fitted, but with Gaussian constraints set according to the external measurements reported in Refs.~\cite{Amhis:2019ckw,Malde:2015mha}. All known correlations are taken in account.  This exercise returns $\dkp = \left( 187.6^{+10.4}_{-11.6} \right)\si\degree$ with a fit quality of $\chi^2 / {\rm n.d.f.} = 0.9/3$.  In order to estimate the relative contributions of the statistical and systematic uncertainties to this result, the fit is re-performed taking only the statistical component of the uncertainties on the measured observables. A comparison of the result from this fit to that of the default procedure leads to the conclusion that the the statistical uncertainty is $\left(^{+8.9}_{-9.7}\right)\si\degree$ and the systematic uncertainty is $\left(^{+5.4}_{-6.4}\right)\si\degree$.

Other fit configurations are investigated, the results of which are presented in Table~\ref{tab:dkpfit}. Removing the external constraints on $\rkp$ and $y$ degrades the sensitivity by around 40\%; on the contrary, fixing these parameters to the central values of the external measurements leads to negligible change in the result.  When taking only $\rkp \sin \dkp$ as input the sensitivity degrades by around 30\%, indicating that the observables sensitive to $\rkp \cos \dkp$ make a valuable contribution to the default result.

\begingroup
\renewcommand{\arraystretch}{1.3} 
\begin{table}[!ht]
    \caption{Results for $\dkp$ for different fit configurations as described in the text.}
    \centering
    \begin{tabular}{lc}\toprule
    Configuration         &  $\dkp$ [$\si\degree$]  \\ \midrule
    Default               & $187.6^{+10.4}_{-11.6}$ \\
    $\rkp$ and $y$ free   & $191.1^{+13.6}_{-14.4}$ \\
    $\rkp$ and $y$ fixed  & $187.6^{+10.4}_{-11.6}$ \\
    $\rkp \sin\dkp$ alone & $190.8^{+14.7}_{-14.0}$ \\ \bottomrule
    \end{tabular}
    \label{tab:dkpfit}
\end{table}
\endgroup

\section{Summary and outlook}
\label{sec:summary}

A double-tag strategy has been employed to determine  the branching fraction of three $D \to K^0_LX$  decays, yielding the results
\begin{eqnarray}
{\cal B}({D^0 \to K^0_L \pi^0}) & = & (0.97 \pm 0.03 \pm 0.02)\,\% \nonumber \\
{\cal B}({D^0 \to K^0_L \omega }) & = &  (1.09 \pm 0.06 \pm 0.03)\,\% \nonumber \\
{\cal B}({D^0 \to K_L^0 \pi^0\pi^0}) & = & (1.26 \pm 0.05 \pm 0.03)\,\% ,  \nonumber 
\end{eqnarray}
\noindent where the first uncertainty is statistical, and the second systematic.   These measurements are the most precise yet performed that are independent of any uncertainty associated with the knowledge of strong-phase parameters, making them valuable inputs for studies of such quantities. 

Using a wide ensemble of tagging modes, including 
these $D \to K^0_LX$  decays, an updated measurement has been performed of $\Akp$, the asymmetry between $C\!P$-odd and $C\!P$-even  $D$-meson decays into $K^-\pi^+$.  In addition, for the first time, a determination has been made of ${\cal A}^{\pi\pi\pi^0}_{K\pi}$, the asymmetry between $D \to K^-\pi^+$ decays tagged with $C\!P$-odd eigenstate modes and the predominantly $C\!P$-even decay $D\to\pi^+\pi^-\pi^0$.   The following values are obtained:
\begin{eqnarray}
{\cal A}_{K\pi} & = & 0.132 \pm 0.011 \pm 0.007 \, \nonumber \\
{\cal A}_{K\pi}^{\pi\pi\pi^0} &= &  0.130 \pm 0.012 \pm 0.008 \, . \nonumber
\end{eqnarray}
The result for ${\cal A}_{K\pi}$ supersedes that reported in Ref.~\cite{Ablikim:2014gvw}, and is around 30\% more precise. 
Both of these observables are sensitive to $\rkp \cos \dkp$.

These asymmetry measurements have been complemented by a study of events containing both $D \to K^-\pi^+$ and $D \to K^0_{S,L}\pi^+\pi^-$ decays, in which the distributions of the three-body modes across their phase spaces are sensitive to both  $\rkp \cos \dkp$ and  $\rkp \sin \dkp$.  A fit to these distributions, together with the asymmetry measurements, gives 
\begin{eqnarray}
    \dkp = \left( 187.6 {^{+8.9}_{-9.7}}{^{+5.4}_{-6.4}} \right)\si\degree \, , \nonumber
\end{eqnarray}
where $\rkp$, $y$ and $F^{\pi\pi\pi^0}_+$ have been constrained to their externally measured values.  This result, which is the most precise to be obtained from quantum-correlated $D\bar{D}$ data, is compatible with that from a global analysis of charm-mixing measurements~\cite{Amhis:2019ckw} and has a similar uncertainty.  It is also consistent with the value determined  from the fit to the LHCb $b$-decay and charm-mixing studies~\cite{LHCb:2021dcr}, and with the prediction made from the phenomenological analysis of two-body charm-meson decay observables~\cite{Buccella:2019kpn}, but has lower precision than both.  However, over the coming few years it is expected that BESIII will accumulate substantially larger data samples at the $\psi(3770)$ resonance~\cite{BESIII:2020nme}, which will allow the sensitivity of the $\dkp$ measurement to be significantly improved.

\section*{Acknowledgments}
The BESIII collaboration thanks the staff of BEPCII and the IHEP computing center for their strong support. This work is supported in part by National Key R$\&$D Program of China under Contracts Nos. 2020YFA0406400, 2020YFA0406300; National Natural Science Foundation of China (NSFC) under Contracts Nos. 11635010, 11735014, 11835012, 11935015, 11935016, 11935018, 11961141012, 12022510, 12025502, 12035009, 12035013, 12192260, 12192261, 12192262, 12192263, 12192264, 12192265; the Chinese Academy of Sciences (CAS) Large-Scale Scientific Facility Program; Joint Large-Scale Scientific Facility Funds of the NSFC and CAS under Contract No. U1832207; CAS Key Research Program of Frontier Sciences under Contract No. QYZDJ-SSW-SLH040; 100 Talents Program of CAS; INPAC and Shanghai Key Laboratory for Particle Physics and Cosmology; ERC under Contract No. 758462; European Union's Horizon 2020 research and innovation programme under Marie Sklodowska-Curie grant agreement under Contract No. 894790; German Research Foundation DFG under Contracts Nos. 443159800, Collaborative Research Center CRC 1044, GRK 2149; Istituto Nazionale di Fisica Nucleare, Italy; Ministry of Development of Turkey under Contract No. DPT2006K-120470; National Science and Technology fund; National Science Research and Innovation Fund (NSRF) via the Program Management Unit for Human Resources \& Institutional Development, Research and Innovation under Contract No. B16F640076; STFC (United Kingdom); Suranaree University of Technology (SUT), Thailand Science Research and Innovation (TSRI), and National Science Research and Innovation Fund (NSRF) under Contract No. 160355; The Royal Society, UK under Contracts Nos. DH140054, DH160214; The Swedish Research Council; U. S. Department of Energy under Contract No. DE-FG02-05ER41374

\bibliographystyle{BESIII}
\bibliography{dkpreferences}

\providecommand{\href}[2]{#2}\begingroup\raggedright\begin{thebibliography}{10}

\bibitem{Aaij:2012nva}
{ {LHCb}} collaboration, \emph{{Observation of $D^0$-$\bar{D}{}^0$
  oscillations}},
  \href{https://doi.org/10.1103/PhysRevLett.110.101802}{\emph{Phys. Rev. Lett.}
  {\bfseries 110} (2013) 101802}
  [\href{https://arxiv.org/abs/1211.1230}{{\ttfamily arXiv:1211.1230}}].

\bibitem{Aaij:2017urz}
{ {LHCb}} collaboration, \emph{{Updated determination of $D^0$-$\bar{D}{}^0$
  mixing and CP violation parameters with $D^0\to K^+\pi^-$ decays}},
  \href{https://doi.org/10.1103/PhysRevD.97.031101}{\emph{Phys. Rev. D}
  {\bfseries 97} (2018) 031101}
  [\href{https://arxiv.org/abs/1712.03220}{{\ttfamily arXiv:1712.03220}}].

\bibitem{Aaltonen:2013pja}
{ CDF} collaboration, \emph{{Observation of $D^0$-$\bar{D}{}^0$ mixing using
  the CDF II Detector}},
  \href{https://doi.org/10.1103/PhysRevLett.111.231802}{\emph{Phys. Rev. Lett.}
  {\bfseries 111} (2013) 231802}
  [\href{https://arxiv.org/abs/1309.4078}{{\ttfamily arXiv:1309.4078}}].

\bibitem{ADS0}
D.~Atwood, I.~Dunietz and A.~Soni, \emph{{Enhanced $CP$ violation with $B \to
  K\Dz(\Dzbar)$ modes and extraction of the Cabibbo--Kobayashi--Maskawa angle
  $\gamma$}}, \href{https://doi.org/10.1103/PhysRevLett.78.3257}{\emph{Phys.
  Rev. Lett.} {\bfseries 78} (1997) 3257}
  [\href{https://arxiv.org/abs/hep-ph/9612433}{{\ttfamily hep-ph/9612433}}].

\bibitem{ADS}
D.~Atwood, I.~Dunietz and A.~Soni, \emph{{Improved methods for observing $CP$
  violation in $B^{\pm}\to K D$ and measuring the CKM phase $\gamma$}},
  \href{https://doi.org/10.1103/PhysRevD.63.036005}{\emph{Phys. Rev. D}
  {\bfseries 63} (2001) 036005}
  [\href{https://arxiv.org/abs/hep-ph/0008090}{{\ttfamily hep-ph/0008090}}].

\bibitem{LHCb:2019hro}
{ {LHCb}} collaboration, \emph{{Observation of $CP$ violation in charm
  decays}}, \href{https://doi.org/10.1103/PhysRevLett.122.211803}{\emph{Phys.
  Rev. Lett.} {\bfseries 122} (2019) 211803}
  [\href{https://arxiv.org/abs/1903.08726}{{\ttfamily arXiv:1903.08726}}].

\bibitem{Ablikim:2014gvw}
{ BESIII} collaboration, \emph{{Measurement of the $D\to K^-\pi^+$ strong phase
  difference in $\psi(3770)\to D^0\bar{D}{}^0$}},
  \href{https://doi.org/10.1016/j.physletb.2014.05.071}{\emph{Phys. Lett. B}
  {\bfseries 734} (2014) 227}
  [\href{https://arxiv.org/abs/1404.4691}{{\ttfamily arXiv:1404.4691}}].

\bibitem{Amhis:2019ckw}
{ HFLAV} collaboration, \emph{{Averages of $b$-hadron, $c$-hadron, and
  $\tau$-lepton properties as of 2018}},
  \href{https://doi.org/10.1140/epjc/s10052-020-8156-7}{\emph{Eur. Phys. J. C}
  {\bfseries 81} (2021) 226}
  [\href{https://arxiv.org/abs/1909.12524}{{\ttfamily arXiv:1909.12524}}],
  {August 2022 update (updated results and plots available at
  \href{https://hflav.web.cern.ch/}{{\texttt{https://hflav.web.cern.ch/}}})}.

\bibitem{LHCb:2020hdx}
{ {LHCb}} collaboration, \emph{{Measurement of $C\!P$ observables in $B^\pm \to
  D^{(*)} K^\pm$ and $B^\pm \to D^{(*)} \pi^\pm$ decays using two-body $D$
  final states}}, \href{https://doi.org/10.1007/JHEP04(2021)081}{\emph{JHEP}
  {\bfseries 04} (2021) 081}
  [\href{https://arxiv.org/abs/2012.09903}{{\ttfamily arXiv:2012.09903}}].

\bibitem{LHCb:2021dcr}
{ {LHCb}} collaboration, \emph{{Simultaneous determination of CKM angle
  $\gamma$ and charm mixing parameters}},
  \href{https://doi.org/10.1007/JHEP12(2021)141}{\emph{JHEP} {\bfseries 12}
  (2021) 141} [\href{https://arxiv.org/abs/2110.02350}{{\ttfamily
  arXiv:2110.02350}}].

\bibitem{Lenz:2020awd}
A.~Lenz and G.~Wilkinson, \emph{{Mixing and CP Violation in the Charm System}},
  \href{https://doi.org/10.1146/annurev-nucl-102419-124613}{\emph{Ann. Rev.
  Nucl. Part. Sci.} {\bfseries 71} (2021) 59}
  [\href{https://arxiv.org/abs/2011.04443}{{\ttfamily arXiv:2011.04443}}].

\bibitem{Buccella:2019kpn}
F.~Buccella, A.~Paul and P.~Santorelli, \emph{{$SU(3)_F$ breaking through final
  state interactions and $CP$ asymmetries in $D \to PP$ decays}},
  \href{https://doi.org/10.1103/PhysRevD.99.113001}{\emph{Phys. Rev. D}
  {\bfseries 99} (2019) 113001}
  [\href{https://arxiv.org/abs/1902.05564}{{\ttfamily arXiv:1902.05564}}].

\bibitem{Ablikim:2021cqw}
{ BESIII} collaboration, \emph{{Measurement of the $D \to K^-\pi^+\pi^+\pi^-$
  and $D \to K^-\pi^+\pi^0$ coherence factors and average strong-phase
  differences in quantum-correlated ${D\bar{D}}$ decays}},
  \href{https://doi.org/10.1007/JHEP05(2021)164}{\emph{JHEP} {\bfseries 05}
  (2021) 164} [\href{https://arxiv.org/abs/2103.05988}{{\ttfamily
  arXiv:2103.05988}}].

\bibitem{Ablikim:2020yif}
{ BESIII} collaboration, \emph{{Determination of strong-phase parameters in
  $D\rightarrow K^0_{S,L}\pi^+\pi^-$}},
  \href{https://doi.org/10.1103/PhysRevLett.124.241802}{\emph{Phys. Rev. Lett.}
  {\bfseries 124} (2020) 241802}
  [\href{https://arxiv.org/abs/2002.12791}{{\ttfamily arXiv:2002.12791}}].

\bibitem{Ablikim:2020lpk}
{ BESIII} collaboration, \emph{{Model-independent determination of the relative
  strong-phase difference between $D^0$ and $\bar{D}^0\rightarrow
  K^0_{S,L}\pi^+\pi^-$ and its impact on the measurement of the CKM angle
  $\gamma/\phi_3$}},
  \href{https://doi.org/10.1103/PhysRevD.101.112002}{\emph{Phys. Rev. D}
  {\bfseries 101} (2020) 112002}
  [\href{https://arxiv.org/abs/2003.00091}{{\ttfamily arXiv:2003.00091}}].

\bibitem{Ablikim:2009aa}
{ BESIII} collaboration, \emph{{Design and construction of the BESIII
  detector}}, \href{https://doi.org/10.1016/j.nima.2009.12.050}{\emph{Nucl.
  Instrum. Meth. A} {\bfseries 614} (2010) 345}
  [\href{https://arxiv.org/abs/0911.4960}{{\ttfamily arXiv:0911.4960}}].

\bibitem{Yu:2016cof}
C.~Yu et~al., \emph{{BEPCII performance and beam dynamics studies on
  luminosity}},  in \emph{{7th International Particle Accelerator Conference}},
  2016, \href{https://doi.org/10.18429/JACoW-IPAC2016-TUYA01}{DOI}.

\bibitem{geant4}
{ GEANT4} collaboration, \emph{{GEANT4 -- a simulation toolkit}},
  \href{https://doi.org/10.1016/S0168-9002(03)01368-8}{\emph{Nucl. Instrum.
  Meth. A} {\bfseries 506} (2003) 250}.

\bibitem{ref:kkmc}
S.~Jadach, B.~Ward and Z.~Was, \emph{{The precision Monte Carlo event generator
  KK for two fermion final states in $e^+e^-$ collisions}},
  \href{https://doi.org/10.1016/S0010-4655(00)00048-5}{\emph{Comput. Phys.
  Commun.} {\bfseries 130} (2000) 260}
  [\href{https://arxiv.org/abs/hep-ph/9912214}{{\ttfamily hep-ph/9912214}}].

\bibitem{Lange:2001uf}
D.~Lange, \emph{{The EvtGen particle decay simulation package}},
  \href{https://doi.org/10.1016/S0168-9002(01)00089-4}{\emph{Nucl. Instrum.
  Meth. A} {\bfseries 462} (2001) 152}.

\bibitem{pdg}
{ Particle Data Group} collaboration, \emph{{Review of particle physics}},
  \href{https://doi.org/10.1093/ptep/ptaa104}{\emph{PTEP} {\bfseries 2020}
  (2020) 083C01}.

\bibitem{PhysRevD.62.034003}
J.~C. Chen, G.~S. Huang, X.~R. Qi, D.~H. Zhang and Y.~S. Zhu, \emph{{Event
  generator for $J/\ensuremath{\psi}$ and $\ensuremath{\psi}(2S)$ decay}},
  \href{https://doi.org/10.1103/PhysRevD.62.034003}{\emph{Phys. Rev. D}
  {\bfseries 62} (2000) 034003}.

\bibitem{YANGRui-Ling:61301}
R.-L. Yang, R.-G. Ping and H.~Chen, \emph{{Tuning and validation of the
  Lundcharm model with $J/\psi$ decays}},
  \href{https://doi.org/10.1088/0256-307X/31/6/061301}{\emph{Chin. Phys. Lett.}
  {\bfseries 31} (2014) 061301}.

\bibitem{RICHTERWAS1993163}
E.~Richter-Was, \emph{{QED bremsstrahlung in semileptonic B and leptonic $\tau$
  decays}},
  \href{https://doi.org/https://doi.org/10.1016/0370-2693(93)90062-M}{\emph{Phys.
  Lett. B} {\bfseries 303} (1993) 163 }.

\bibitem{He:2007aj}
{ CLEO} collaboration, \emph{{Comparison of $D \to K^0_S \pi$ and $D \to K^0_L
  \pi$ decay rates}},
  \href{https://doi.org/10.1103/PhysRevLett.100.091801}{\emph{Phys. Rev. Lett.}
  {\bfseries 100} (2008) 091801}
  [\href{https://arxiv.org/abs/0711.1463}{{\ttfamily arXiv:0711.1463}}].

\bibitem{Asner:2012xb}
{ CLEO} collaboration, \emph{{Updated measurement of the strong phase in $D^0
  \to K^+\pi^-$ decay using quantum correlations in $e^+e^- \to D^0 \bar{D}^0$
  at CLEO}}, \href{https://doi.org/10.1103/PhysRevD.86.112001}{\emph{Phys. Rev.
  D} {\bfseries 86} (2012) 112001}
  [\href{https://arxiv.org/abs/1210.0939}{{\ttfamily arXiv:1210.0939}}].

\bibitem{BESIII:2022xhe}
{ BESIII} collaboration, \emph{{Measurements of absolute branching fractions of
  $D^0\to K_L^0\phi$, $K_L^0\eta$, $K_L^0\omega$, and $K_L^0\eta^{\prime}$}},
  \href{https://doi.org/10.1103/PhysRevD.105.092010}{\emph{Phys. Rev. D}
  {\bfseries 105} (2022) 092010}
  [\href{https://arxiv.org/abs/2202.13601}{{\ttfamily arXiv:2202.13601}}].

\bibitem{Malde:2015mha}
S.~Malde, C.~Thomas, G.~Wilkinson, P.~Naik, C.~Prouve, J.~Rademacker et~al.,
  \emph{{First determination of the $C\!P$ content of $D \to
  \pi^+\pi^-\pi^+\pi^-$ and updated determination of the $CP$ contents of $D
  \to \pi^+\pi^-\pi^0$ and $D \to K^+K^-\pi^0$}},
  \href{https://doi.org/10.1016/j.physletb.2015.05.043}{\emph{Phys. Lett. B}
  {\bfseries 747} (2015) 9} [\href{https://arxiv.org/abs/1504.05878}{{\ttfamily
  arXiv:1504.05878}}].

\bibitem{JohsnsonSU}
N.~L. Johnson, \emph{Systems of frequency curves generated by methods of
  translation}, {\emph{Biometrika} {\bfseries 36} (1949) 149}.

\bibitem{BESIII:2018apz}
{ BESIII} collaboration, \emph{{Measurements of absolute branching fractions
  for $D$ mesons decays into two pseudoscalar mesons}},
  \href{https://doi.org/10.1103/PhysRevD.97.072004}{\emph{Phys. Rev. D}
  {\bfseries 97} (2018) 072004}
  [\href{https://arxiv.org/abs/1802.03119}{{\ttfamily arXiv:1802.03119}}].

\bibitem{BESIII:2018iev}
{ BESIII} collaboration, \emph{{Measurement of $e^+e^- \to {D} \bar{D}$ cross
  sections at the $\psi(3770)$ resonance}},
  \href{https://doi.org/10.1088/1674-1137/42/8/083001}{\emph{Chin. Phys. C}
  {\bfseries 42} (2018) 083001}
  [\href{https://arxiv.org/abs/1803.06293}{{\ttfamily arXiv:1803.06293}}].

\bibitem{ALBRECHT1990278}
{ ARGUS} collaboration, \emph{Search for hadronic $b\to u$ decays},
  \href{https://doi.org/https://doi.org/10.1016/0370-2693(90)91293-K}{\emph{Phys.
  Lett. B} {\bfseries 241} (1990) 278}.

\bibitem{Libby:2010nu}
{ CLEO} collaboration, \emph{{Model-independent determination of the
  strong-phase difference between $D^0$ and $\bar{D}^0 \to K^0_{S,L} h^+ h^-$
  ($h=\pi,K$) and its impact on the measurement of the CKM angle
  $\gamma/\phi_3$}},
  \href{https://doi.org/10.1103/PhysRevD.82.112006}{\emph{Phys. Rev. D}
  {\bfseries 82} (2010) 112006}
  [\href{https://arxiv.org/abs/1010.2817}{{\ttfamily arXiv:1010.2817}}].

\bibitem{Aubert:2008bd}
{ BaBar} collaboration, \emph{{Improved measurement of the CKM angle $\gamma$
  in $B^\mp \to D^{(*)} K^{(*\mp)}$ decays with a Dalitz plot analysis of $D$
  decays to $K^0_{S} \pi^{+} \pi^{-}$ and $K^0_{S} K^{+} K^{-}$}},
  \href{https://doi.org/10.1103/PhysRevD.78.034023}{\emph{Phys. Rev. D}
  {\bfseries 78} (2008) 034023}
  [\href{https://arxiv.org/abs/0804.2089}{{\ttfamily arXiv:0804.2089}}].

\bibitem{BESIII:2020nme}
{ BESIII} collaboration, \emph{{Future physics programme of BESIII}},
  \href{https://doi.org/10.1088/1674-1137/44/4/040001}{\emph{Chin. Phys. C}
  {\bfseries 44} (2020) 040001}
  [\href{https://arxiv.org/abs/1912.05983}{{\ttfamily arXiv:1912.05983}}].

\end{thebibliography}\endgroup

\end{document}